\documentclass[journal=jctcce,manuscript=perspective,layout=twocolumn]{achemso}
\pdfoutput=1
\usepackage[version=3]{mhchem} 
\usepackage{color,dsfont,bm}
\usepackage{float,graphicx,subfig}
\newcommand*{\szero}{|0,0\rangle}

\newcommand*{\tup}{|1,+\rangle}
\newcommand*{\tdown}{|1,-\rangle}
\newcommand*{\tzero}{|1,0\rangle}

\author{Jason K. Ellis}
\author{Richard L. Martin}
    \email{rlmartin@lanl.gov}
    \affiliation[Los Alamos National Laboratory]{Theoretical Division, Los Alamos National Laboratory, Los Alamos, NM 87545}

\author{Gustavo E. Scuseria}
    \affiliation[Rice University]{Department of Chemistry and Department of Physics and Astronomy, Rice University, Houston, TX 77005-1827}

\title[Pair Functions]{On pair functions for strong correlations} 

\begin{document}
\begin{abstract}
The UHF wave function may be written as a spin-contaminated \textit{pair} wave function of the APSG form, and the overlap of the alpha and beta corresponding orbitals of the UHF solution can be taken as a proxy for the strength of the correlation captured by breaking symmetry.  We demonstrate this with calculations on one- and two-dimensional hydrogen clusters and make contact with the well studied Hubbard model. The UHF corresponding orbitals pair in a manner that allows a smooth evolution from doubly occupied orbitals at small distance to one in which wave function breaks symmetry, segregating the $\alpha$ and $\beta$ electrons onto distinct sublattices at large distances. By performing spin projection on these UHF solutions, we address strong correlations that are difficult to capture at intermediate distances using a single determinant.  Approved for public release: LA-UR-13-22691.
\end{abstract}

\section{Introduction}
\label{sec:Intro}
The proper treatment of what has come to be known as the ``strong correlation'' problem remains a fundamental challenge to our understanding of unconventional superconductivity, frustrated spin lattices, the heavy fermion problem, and the class of materials known as Mott insulators. Materials are typically considered strongly correlated when conventional density functional approximations fail to qualitatively describe their properties. For example, the local density and generalized gradient approximations (the LDA and GGA) of density functional theory (DFT) predicted NiO to be a ferromagnetic metal. This catastrophically failed to describe the experimental properties: NiO behaves as an antiferromagnetic insulator with an optical gap of the order of 4 eV. \cite{2002PhRvB..65o5102D}  

This led to statements by some that ``DFT cannot describe Mott insulators.'' Recent research has shown that this failure lies in the local or semi-local exchange-correlation approximations typically employed in the field, and not with DFT itself. Including a fraction of the fully \textit{non-local} HF exchange term in the functional remedies many of these problems, and these hybrid functionals yield significantly improved band gaps, lattice constants, and magnetic properties for Mott insulators.\cite{Wen:2012fp,2011PSSBR.248..767H} The strong correlations responsible for opening a gap in NiO do not vanish if it is made metallic by application of pressure, and so conventional local and semi-local approximations to DFT have problems with strongly correlated metals as well. Mounting evidence suggests that the hybrid functionals do not perform well in this regime either. \cite{2008JPCM...20f4201M,2007JChPh.127b4103P,Wen:2012fp} 

Theorists have struggled with this problem for decades, and many approaches that build upon traditional DFT have attempted to address it. These include the self-interaction-correction (SIC) to DFT,\cite{1981PhRvB..23.5048P} DFT+U approaches,\cite{1991PhRvB..44..943A,2000:anisimov+LDA+U} and many-body approaches such as the GW approximation,\cite{Hedin:1965hi,Hedin19701} and dynamic mean-field theory (DMFT).\cite{1996RvMP...68...13G} The SIC offered a significant improvement over conventional DFT in the Mott insulator regime but vanishes for a metal, necessitating additional approximations in that realm.\cite{2010PhRvB..81d5108P} The GW approach, a many-body Green's function method based on the random phase approximation, successfully predicted band gaps in many materials and some Mott insulators. However, the strength of this method, particularly its focus on the quasiparticle excitation spectrum, also made it difficult to obtain a total energy even though this energy is connected to the RPA ground state correlation.\cite{2000PhRvL..85.5611S} DMFT, while formally a non-empirical embedding method, relies in practice on material dependent parameters for its utilization, as do the DFT+U approaches.\cite{Wen:2012fp} A truly predictive, parameter free, \textit{ab initio} description of the electronic structure of strongly correlated materials remains a significant challenge for theory. 

\begin{figure}[t]
	\centering
	\includegraphics[width=2.9in]{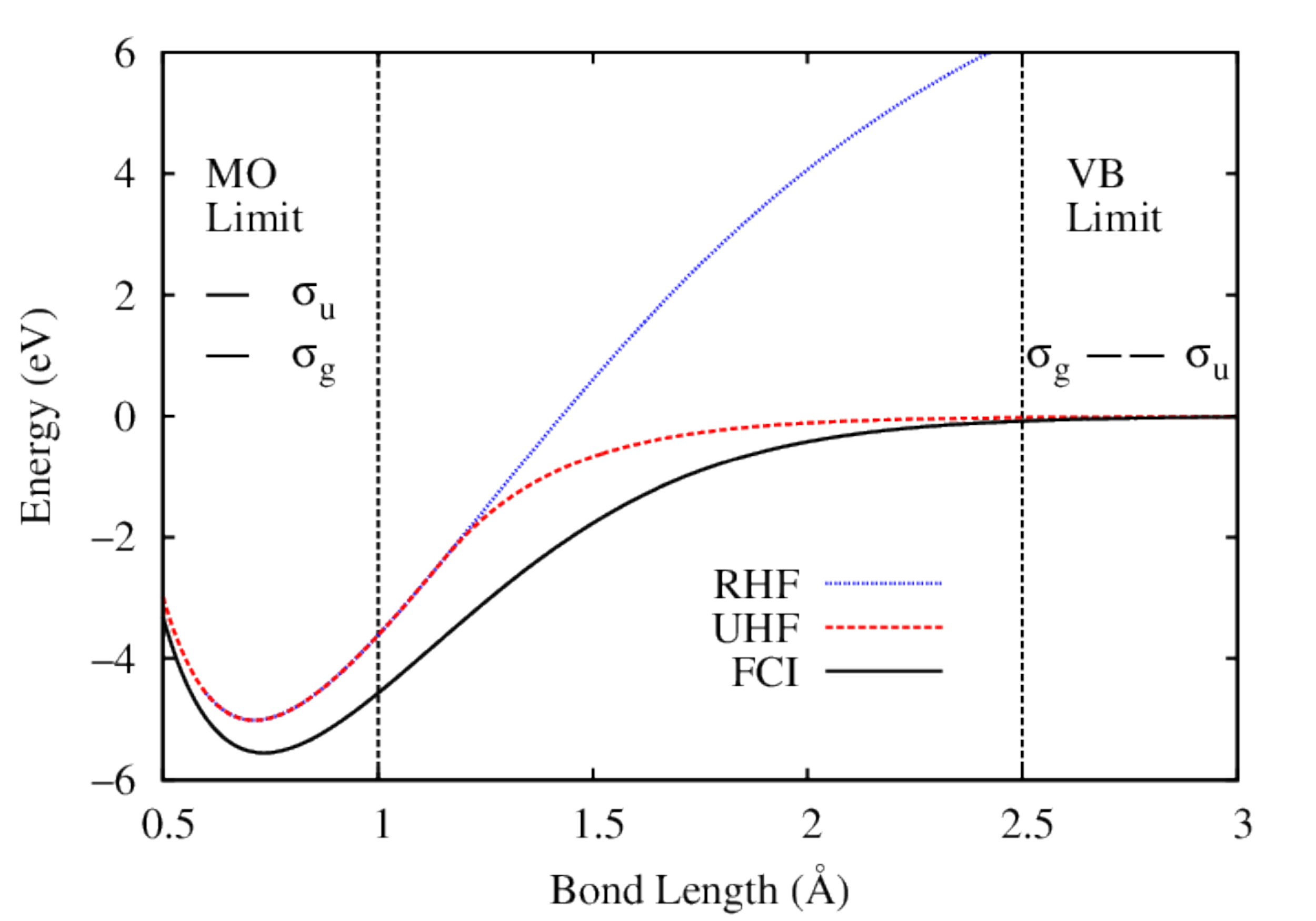}
	\caption{ Dissociation curve for \ce{H2} computed with the STO-3G basis set for illustration purposes. The RHF single determinant qualitatively describes the molecule near equilibrium but does not dissociate correctly (see text). By artificially breaking spatial and spin symmetry, the single determinant UHF solution yields a qualitatively correct curve in both limits.}
	\label{fig:H2}
\end{figure}

The Mott transition bears striking similarities to one of the oldest problems in quantum chemistry: making and breaking chemical bonds. Specifically, consider breaking the simplest bond, that of \ce{H2} (see Figure \ref{fig:H2}). Near equilibrium, a symmetry restricted Hartree-Fock (RHF), doubly occupied $\sigma_{g}$ molecular orbital (MO), dominates the wave function. This description fails abysmally in the dissociation limit; the double occupancy implicit in the $\sigma_{g}^{2}$ configuration contributes energetically unfavorable ionic components to the wave function at large bond lengths. In this limit, the bonding and anti-bonding molecular orbitals become nearly degenerate; when expressed in this symmetry restricted MO basis, a proper wave function requires two configurations, $\sigma_{g}^{2}$ and $\sigma_{u}^{2}$ with approximately equal contributions (the full configuration interaction, or FCI wave function in a minimum basis). A superposition of these two configurations with variationally determined weights evolves smoothly from the delocalized MO limit to a localized valence bond (VB) description.  Similarly, the transition from a simple metal to a typical Mott insulator requires a wave function which continuously evolves from the independent electron, delocalized Bloch orbital description into a strongly correlated, valence bond-like wave function as the localization increases.  

The success of hybrid functionals in describing Mott insulators relies, in part, on relaxing symmetry constraints imposed on the molecular orbitals. In the unrestricted Hartree-Fock (UHF) or Kohn-Sham (UKS) approximation each electron occupies its own spatial orbital. \cite{Pople:1954gz} This allows the wave function to localize the electrons when appropriate, one to each atom in \ce{H2} at large distance or one per site in an anti-ferromagnet, thereby avoiding the unfavorable ionic component implicit in the restricted approximation. Relaxing symmetry constraints sacrifices the quantum numbers associated with both spatial and spin symmetries. Nevertheless, the unrestricted (UHF) wave function allows a qualitatively (if not quantitatively) acceptable description of both the Mott insulator and simple metal limits. We should not be surprised that hybrid functionals have difficulty describing the correlated metal regime,\cite{2008JPCM...20f4201M,2011PSSBR.248..767H,2007JChPh.127b4103P,Wen:2012fp} corresponding to the intermediate bond lengths of Figure \ref{fig:H2}. In this region, strong correlations expose the problems inherent in single determinant methods, and a successful theory must delicately balance the competition between localization and delocalization.

In this work, we focus on understanding this intermediate ``recoupling'' region, in which neither the MO nor VB picture, neither the restricted or unrestricted approximations are appropriate. We begin by reviewing some concepts regarding broken symmetry and pair wave functions, then briefly discuss wave functions which cannot be expressed in terms of pair functions. Next we describe projection methods which restore the symmetries, illustrated with calculations on finite size hydrogen networks. Finally, we conclude by discussing the implications for truly periodic networks.

\section{Pair Wave Functions}\label{sec:Theory}

We begin by reviewing the wave function of diatomic hydrogen as a function of bond distance. Our discussion utilizes a minimal basis set, the simplest expansion which captures the near degeneracies between the molecular orbitals that occur as the bond is stretched. A minimum basis set also provides a conceptual link with the Hilbert space utilized in the widely studied Hubbard model,\cite{Hubbard:1963uf} though we retain the exact Hamiltonian in the calculations to follow. These near degeneracies give rise to our primary concern, the so-called left-right, or static correlation. Static correlation influences the \textit{form} of the wave function, and can generally be accounted for in the Hilbert space of the minimum basis. In a condensed matter context this might be referred to as antiferromagnetic correlation. Dynamic correlation, or the instantaneous avoidance of one electron by another, constitutes a much weaker effect and, generally speaking, the local and semi-local correlation functionals of DFT describe it adequately. That is to say, for a molecule qualitatively well described by a single determinant of doubly occupied orbitals then many-body perturbation theory, coupled-cluster theory, etc., can describe the dynamic correlation, although it requires an extension of the basis set.

To this end, consider a single $1s$ orbital centered on each of two hydrogen atoms, denoting these as $\phi_{l}$ and $\phi_{r}$, assigned to the left and right atomic centers, respectively. For simplicity, we assume orthogonality of these atomic wave functions, though this is not necessary in principle. Form molecular orbitals using the standard linear combinations of atomic orbitals, giving gerade ($\sigma_{g}$, bonding) and ungerade ($\sigma_{u}$, anti-bonding) orbitals:
\begin{align}
	\sigma_{g} = \frac{\phi_{l} + \phi_{r}}{\sqrt{2}} && \text{and} &&\sigma_{u} = \frac{\phi_{l} - \phi_{r}}{\sqrt{2}}.
\end{align}
The RHF determinant with the bonding orbital doubly occupied approximates the exact wave function quite closely at the equilibrium bond length. This wave function can be written as:
\begin{equation}\label{eq:RHF}
	\mathcal{A}\left[\sigma_{g}(1)\alpha(1)\sigma_{g}(2)\beta(2)\right]=\mathcal{A}[\sigma_{g}\overline{\sigma}_{g}],
\end{equation}
where $\mathcal{A}$ is the anti-symmetrizer, $\alpha$ and $\beta$ are the normal spin functions and we have used an overline to indicate beta spin-orbitals. One may then expand this wave function in the original atomic orbital basis:
\begin{equation}
	\sigma_{g}\overline{\sigma}_{g} = |\Phi_{g}^{C}\rangle \szero +  |\Phi_{g}^{I}\rangle \szero,
\end{equation}
by defining the following spatial functions:
\begin{subequations}
\begin{align}
|\Phi_{g}^{C}\rangle & \equiv \frac{1}{\sqrt{2}}\left(\phi_{l} \phi_{r} + \phi_{r}, \phi_{l}\right)\\
|\Phi_{g}^{I}\rangle & \equiv \frac{1}{\sqrt{2}}\left(\phi_{l} \phi_{l} + \phi_{r} \phi_{r}\right),
\end{align} 
\end{subequations}
and the singlet spin wave function:
\begin{equation}
	|0,0\rangle =  \frac{\left(\alpha\beta - \beta\alpha\right)}{\sqrt{2}}.
\end{equation}
These two spatial functions describe the two components of gerade symmetry that contribute to the $^1\Sigma_{g}$ ground state of \ce{H2}. The first term, $|\Phi_{g}^{C}\rangle$, describes the covalent portion of the wave function, in the Heitler-London (HL) sense. The ionic component, $|\Phi_{g}^{I}\rangle$, describes the symmetric combination ($\Sigma_{g}$) of charge-transfer terms \ce{H+}\ce{H-} and \ce{H-}\ce{H+}. For completeness, define two spatial wave functions of ungerade symmetry that play roles in the excited $^1\Sigma_{u}$ and $^3\Sigma_{u}$ states and appear later in the discussion of the UHF wave function:
\begin{subequations}
\begin{align}
|\Phi_{u}^{C}\rangle & \equiv \frac{1}{\sqrt{2}}\left(\phi_{l} \phi_{r} - \phi_{r} \phi_{l}\right)\label{eq:ungerade}\\
|\Phi_{u}^{I}\rangle & \equiv \frac{1}{\sqrt{2}}\left(\phi_{l} \phi_{l} - \phi_{r} \phi_{r}\right).
\end{align} 
\end{subequations}

At long range, the correct wave function for the ground state of \ce{H2} contains only the HL term, $|\Phi_{g}^{C}\rangle$. The RHF wave function dissociates improperly precisely because the ionic piece imposes an energy cost equal to the ionization potential of one hydrogen atom minus the electron affinity of the other. This quantity is referred to as $U$ in the Hubbard ubiquitious in the condensed matter literature and $\gamma_{ii}$ in the Pariser-Parr-Pople model familiar to quantum chemists. It is a simple measure of the on-site electron-electron repulsion.

Qualitatively, what does this mean? At small distances, delocalizing the electrons over both centers allows the electrons to gain sufficient kinetic energy to overcome the on-site repulsion. At large distances, the kinetic energy obtained from delocalization becomes small, overwhelmed by the associated on-site electron repulsion, and the electrons localize. An acceptable approximation requires a wave function with can evolve smoothly between these two limits.

\subsection{The Generalized Valence Bond Wave Function}
The simplest wave function for \ce{H2} that smoothly evolves from a delocalized MO description to a localized VB description is the generalized valence bond (GVB) wave function. \cite{Hurley:1953wh,Hunt:1972bh,MOSS:1975us,Bobrowicz:1977ta} For two electrons this may be thought of as a multi-determinant extension of RHF to an open-shell singlet pair (geminal) wave function of the form:
 \begin{equation}
 \label{eq:gvb}
 \Psi_{GVB} =\mathcal{A} \left[\chi_l \chi_r\right] \szero,
 \end{equation}
where the \textit{non-orthogonal} spatial functions $\chi_{i}$ must be determined self-consistently at each distance. The non-orthogonality allows each orbital to polarize continuously as a function of bond length, providing enough variational flexibility to successfully break the bond.

Expanding each of these non-orthogonal orbitals in the original MO basis:  
\begin{align}\label{eq:mixorb}
         \chi_l &= \frac{\sigma_{g} + \lambda^{1/2} \sigma_{u}}{\sqrt{(1+\lambda)}} && \text{and}&
	\chi_r &= \frac{\sigma_{g} - \lambda^{1/2} \sigma_{u}}{\sqrt{(1+\lambda)}}
\end{align}
introduces a mixing parameter $\lambda$, a real number between -1 and 1. The spatial overlap of the non-orthogonal orbitals can be calculated:
\begin{equation}
	S_{lr} = \left\langle\chi_l|\chi_r\right\rangle = \frac{1-\lambda}{1+\lambda},
\end{equation}
and after some algebra, Eq.\eqref{eq:gvb} becomes:
\begin{equation}
\begin{split}
	\Psi_{UHF} &= \frac{1+\lambda}{\sqrt{2}(1+\lambda^{2})} |\Phi_{g}^{C}\rangle\szero \\
    &+ \frac{1-\lambda}{\sqrt{2}(1+\lambda^{2})}|\Phi_{g}^{I}\rangle\szero
\end{split}
\end{equation}
This form makes explicit the relationship between the ionic and covalent portions of the wave function, while remaining equivalent to the multi-configuration representation more familiar to quantum chemists:
\begin{equation}
	\frac{1} {\sqrt{1+\lambda^{2}}}\left[\sigma_{g}^{2} - \lambda \sigma_{u}^{2}\right]\szero.
\end{equation}
For a single pair in a minimal basis set, this wave function is exact, with the mixing parameter $\lambda$ and the orbitals determined self-consistently. 

We have belabored this simple two-electron example because we wish to emphasize the usefulness of this construct for strongly correlated materials. Pushing the solid-state analogy further, this simple pair function continuously connects an ionic charge-density-wave $(\lambda = -1)$, a simple metal $(\lambda = 0)$, and the antiferromagnetic singlet $(\lambda = 1)$ with a single mixing parameter. 

To extend this pairing concept to systems with more electrons, one writes the wave function for $N_{e}$ electrons as an anti-symmetrized product of geminals (APG). For an extensive review of geminal wave functions see, for example, the work of \citet{springerlink:10.1007/3-540-48972-X_4} and the references therein. In the APG formalism, electrons not explicitly correlated are grouped in the core and frozen from excitation. Each correlated MO $\psi_i$ pairs with an unoccupied partner $\psi_{i^{\prime}}$ in the following manner:
\begin{equation}\label{eq:PairWfn}
\psi = \mathcal{A}[core]\prod_{i=1}^{N} \frac{\psi_i\overline\psi_i - \lambda_{i}\psi_{i^{\prime}}\overline\psi_{i^{\prime}}}{\sqrt{1+\lambda_{i}^{2}}}.
\end{equation}
If one imposes the condition of strong orthogonality between the pairs, the wave function is denoted as an anti-symmetrized product of strongly orthogonal geminals (APSG). In addition to strong orthogonality, enforced by restricting each pair to distinct occupied and virtual orbital spaces, one may further restrict the pairs to pure spin states, a constraint denoted as ``perfect pairing''.\cite{MOSS:1975us} Such a wave function explicitly neglects inter-pair correlation effects, as well as \textit{same-spin} correlation, that may be important in many cases, particularly for high spin multiplets.

\subsection{Broken Symmetry Wave Functions}
The UHF approach eschews the multi-determinant nature of GVB in favor of a single determinant wave function. Allowing different spins to occupy different spatial orbitals provides enough flexibility for the electrons to balance the contributions from the covalent and ionic terms. At short distances this approach yields the RHF wave function, but as the bond stretches past some critical distance, known as the Coulson-Fischer (CF) point, the kinetic energy and nuclear attraction terms associated with the ionic portion no longer compensate for the on-site repulsion, and a distinct \textit{broken-symmetry} UHF determinant emerges as the ground state. Artificially breaking spatial symmetry by placing electrons of different spin in different orbitals finesses the electron correlations that dissociate the bond, projecting away the ionic component of the RHF wave function. Breaking symmetry in the presence of orbital near degeneracies should not lower the energy, but the fact that the UHF energy lies lower than the RHF energy lies at the heart of L\"{o}wdin's symmetry dilemma: one obtains a better approximation to the total energy by artifactually sacrificing good quantum numbers. \cite{Lykos:1963ep}

Returning to the simple two-electron case, the UHF wave function takes the form:
\begin{equation}
	\Psi_{UHF} = \chi_{l}\bar{\chi}_{r},
\end{equation}
where $\chi_{l}$ and $\chi_{r}$ are the non-orthogonal orbitals defined by the mixing coefficient $\lambda$ in Equation \eqref{eq:mixorb}. In the dissociation limit, these orbitals reduce to the atomic orbitals associated with each atom, and the wave function becomes either:
\begin{align*}
	\phi_{l} \bar{\phi}_{r}&& \text{or} &&\bar{\phi}_{l}\phi_{r},
\end{align*}
both of which break spatial symmetry. The spin products $\alpha\beta$ and $\beta\alpha$ are not eigenfunctions of spin, but rather a linear combination of the singlet and triplet functions: 
\begin{align}
	\alpha\beta &= \frac{|0,0\rangle + |1,0\rangle}{\sqrt{2}} &\text{and}\\ \beta\alpha &= \frac{|1,0\rangle - |0,0\rangle}{\sqrt{2}},
\end{align}
where $|1,0\rangle$ is the $m_{s} = 0$ component of the triplet:
\begin{equation}
	|1,0\rangle =  \frac{\left(\alpha\beta + \beta\alpha\right)}{\sqrt{2}}.
\end{equation}

Rewriting the full UHF wave function by separating the spatial parts into symmetric and anti-symmetric components, we identify the spatial functions associated with the singlet and triplet spin components:
\begin{equation}\label{eq:PsiUHF}
\begin{split}
	\Psi_{UHF} &= \frac{1+\lambda}{\sqrt{2}(1+\lambda)} |\Phi_{g}^{C}\rangle\szero \\
    &+ \frac{1-\lambda}{\sqrt{2}(1+\lambda)}|\Phi_{g}^{I}\rangle\szero \\
    &- \frac{2\lambda}{\sqrt{2}(1+\lambda)}|\Phi_{u}^{C}\tzero.
\end{split}
\end{equation}
The singlet component (up to normalization) is identical to that of the GVB wave function, and the anti-bonding covalent spatial function $|\Phi_{C}^{-}\rangle$ from Equation \eqref{eq:ungerade} possesses ungerade symmetry to couple to the $m_{s} = 0$ triplet spinor. Thus, the variational parameter $\lambda$, which becomes non-zero for all bond lengths greater than the CF point, controls the spin symmetry breaking. In the limit where $\lambda \rightarrow 1$ these orbitals simply reduce to $\phi_{l}$ and $\phi_{r}$ above; in the limit $\lambda \rightarrow 0$ the wave function becomes identical to RHF; and in the limit $\lambda \rightarrow -1$ the wave function describes a spin-contaminated charge density wave. While the UHF wave function correctly dissociates the molecule using a single determinant, this comes at a rather high price: a discontinuity in the derivative of the energy with respect to the internuclear separation occurs at the CF point, and at large distances the wave function no longer possesses the correct parity or spin. 

It is, of course, possible to break further symmetries of a single-determinant wave function. These have been explored in detail by Fukutome\cite{Fukutome:1981du} as well as Stuber and Paldus.\cite{Stuber03ex} In our opinion, this work has not received the attention that it deserves because breaking additional symmetries sacrifices more quantum numbers that, in turn, become quite difficult to recover. An approach of particular note, the generalized Hartree-Fock approximation (GHF), allows each electron to occupy an orbital without definite $\alpha$ or $\beta$ spin. Computationally, one typically accomplishes this by allowing the MO coefficients to become complex, i.e. each real atomic orbital gets both a real and imaginary $\alpha$ and $\beta$ coefficient.  The GHF approach is fraught with other difficulties, as dissociation curves often contain energy crossings and other weird behavior. For these reasons, the community has largely neglected the GHF approach. However, preserving some symmetries of the wave function in a single determinant approach (e.g. UHF) comes with an additional price: closed shell molecules can't in general be dissociated into the correct open-shell fragments. However, using the GHF approach recovers size consistency in the resulting wave function.\cite{JimenezHoyos:2011ia}  We direct the reader to the rather vast literature on the subject of geminal functions and symmetry for an in depth discussion (see, for example, ref. (\citenum{springerlink:10.1007/3-540-48972-X_4}) and (\citenum{Stuber03ex}) and the references therein), and have simply presented a few conceptual ideas here which shed light on some of the issues that arise when we encounter spin projection below. 

\subsection{Pairing in Broken Symmetry Wave Functions}

The UHF wave function possesses a particularly interesting property, pointed out in different forms by many authors. L\"{o}wdin proposed the ``pairing'' theorem\cite{Lowdin:1955zz} which states that the orbitals of a wave function of the UHF type can always be transformed such that the alpha and beta orbitals are mutually orthogonal except for ``corresponding pairs'' of orbitals which have non-zero overlap. This follows from the corresponding orbital transformation of Amos and Hall\cite{1961RSPSA.263..483A} or, alternatively, from the left and right eigenvectors of a singular value decomposition.

Consider a UHF wave function defined by a set of occupied alpha spin orbitals, $\psi^\alpha$, and a set of occupied beta spin orbitals, $\psi^\beta$, orthonormal within themselves, but not mutually so. Denoting overlap between the two sets as $S_{i,j}^{\alpha,\beta}$, they can be brought to bi-orthogonal form by diagonalizing the matrices  $\mathbf{SS}^{\dagger}$ and $\mathbf{S}^{\dagger}\mathbf{S}$ withing the subspace of the occupied $\alpha$ and $\beta$ electrons, respectively, to generate two new sets of spin-orbitals, $\mathbf{U}\psi^\alpha$ and $\mathbf{V}\psi^\beta$: 
\begin{align}
	\mathbf{SS}^{\dagger} \mathbf{U} = \xi \mathbf{U} &&\text{and} &&\mathbf{S}^{\dagger}\mathbf{S V} = \xi^{\prime} \mathbf{V}.
\end{align}
The overlap matrix becomes diagonal in these subspaces, i.e.: 
\begin{equation}
	\mathbf{U^{\dagger} S V} = \delta_{ij}\sqrt{\xi}_{i},
\end{equation}
and the eigenvalues $\xi_{i}=\xi^{\prime}$ are bound between 0 and 1. Thus, each $\alpha$ orbital pairs with a single $\beta$ orbital, and the overlap for pair $i$ is given by $\sqrt{\xi}_{i}$.  

Although a somewhat straightforward result, it does not seem to have been emphasized that because the corresponding orbitals $\mathbf{U}$ and $\mathbf{V}$ are obtained from unitary transformations of the occupied $\alpha$ and $\beta$ UHF orbitals, the UHF wave function may be written as an APSG:
 
\begin{equation}
	\psi = \mathcal{A}[core]\prod_{i=1}^{N} \omega_i(1,2),
\end{equation}
where the geminals $\omega_{i}(1,2)$ are given by:
\begin{equation}\label{eq:uhfpair}
	\omega_i(1,2) = u_i v_i  \alpha\beta, 
\end{equation}
and the spatial overlap of the non-orthogonal alpha and beta orbitals making up the pair is:
\begin{equation}
	S_{ij} = \left\langle u_i|v_j\right\rangle = \sqrt{\xi}_i.
\end{equation}

This pair function extends the GVB pair idea, retaining the condition of strong orthogonality, but relaxing the perfect pairing constraint and allowing spin contamination from the triplet component of the state $\alpha\beta$ associated with each geminal. The non-orthogonal representation of the pair above can of course be transformed back into an orthogonal ``natural orbital'' representation, or expressed in the space of the symmetry restricted molecular orbitals as in Equation \ref{eq:mixorb}. Thus the UHF wave function also contains only intra-pair correlation, though relaxing the perfect pairing constraint allows it the ability to capture more correlation than the GVB approach at the cost of the spin quantum numbers. Finally, we wish to emphasize that the GHF wave function is not an APG, as it also captures inter-pair correlation, retaining size consistency.

\subsection{The Broken Symmetry 1PDM}

L\"{o}wdin pointed out the usefulness of the one-particle density matrix (1PDM) for defining the natural spin-orbitals of a system, and noted that the eigenvalues of this matrix were constrained to be between zero and one.\cite{Lowdin:1955zz} Coleman later argued that all two-particle interactions can be described using the one and two-particle density matrices. \cite{1963RvMP...35..668C} We therefore pause for a moment to remark on the structure of the 1PDM of the UHF and GHF wave functions. 

One can separate the 1PDM into spin blocks and write the 1PDM of a UHF wave function in the following way:
\begin{align}\label{eq:UHF-1pdm}
	\bm{\rho} &= \begin{pmatrix} \bm{\rho}_{\alpha\alpha} & \mathbf{0}\\\mathbf{0} & \bm{\rho}_{\beta\beta}
	\end{pmatrix}\\& = \begin{pmatrix} \mathbf{P}+\mathbf{M}_{z} & \mathbf{0} \\ \mathbf{0} & \mathbf{P} - \mathbf{M}_{z}
	\end{pmatrix}.
\end{align}
The half sum and half difference of the non-zero blocks form the charge ($\mathbf{P}$) and spin densities ($\mathbf{M}_{z}$), respectively:
\begin{align}
	\mathbf{P} &= \frac{\bm{\rho}_{\alpha\alpha} + \bm{\rho}_{\beta\beta}}{2} && \text {and} \\
	\mathbf{M}_{z} &= \frac{\bm{\rho}_{\alpha\alpha} - \bm{\rho}_{\beta\beta}}{2}.
\end{align}
The magnetization describes the UHF spin contamination arising from the triplet contribution to the broken symmetry wave function. We call such a wave function \textit{collinear} because $\hat{S}_{z}$ remains a good quantum number.

The 1PDM associated with the UHF wave function must be both Hermitian and idempotent since this wave function is a single Slater determinant. One can then trivially show the idempotency of each each spin block $\bm{\rho}_{\sigma\sigma}$. Thus, the charge density is a linear combination of two idempotent matrices, and consequently has eigenvalues which are either 0, 1, 1/2, or come in ``corresponding pairs'' of $n$ and $1-n$.\cite{Harriman:1964fr,Rabanovich:2004dd} The natural orbitals of the charge density coupled as ``corresponding pairs'' are simply the natural orbital representation of the non-orthogonal UHF geminals discussed above.

In the GHF formalism, even though additional symmetries are broken, the wave function is still a single Slater determinant though the 1PDM takes the more general form:
\begin{align}
	\bm{\rho} &= \begin{pmatrix} \bm{\rho}_{\alpha\alpha} & \bm{\rho}_{\alpha\beta}\\\bm{\rho}_{\beta\alpha} & \bm{\rho}_{\beta\beta}
	\end{pmatrix}\\ &= \begin{pmatrix} \mathbf{P}+\mathbf{M}_{z} & \mathbf{M}_{x}-i\mathbf{M}_{y} \\ \mathbf{M}_{x}+\mathbf{M}_{y} & \mathbf{P} - \mathbf{M}_{z}
	\end{pmatrix},
\end{align}
where the x and y components of the magnetization are given by:
\begin{align}
	\mathbf{M}_{x} &= \frac{\bm{\rho}_{\alpha\beta} + \bm{\rho}_{\beta\alpha}}{2} && \text{and} \\
	\mathbf{M}_{y} &= \frac{\bm{\rho}_{\alpha\beta} - \bm{\rho}_{\beta\alpha}}{2i}.
\end{align}
For a GHF wave function, no orbital rotation can bring the 1PDM to the form of Equation \eqref{eq:UHF-1pdm}, and we call such a state \textit{non-collinear}, as the state no longer has a good quantum number associated with $\hat{S}_{z}$.

The most general GHF wave function cannot be written as the sum of two idempotent matrices, and loses the property of corresponding pairs. This results from the fact that the wave function cannot be written as an APG. Where the UHF formalism can only describe intra-pair correlation, the GHF formalism has additional flexibility to capture inter-pair correlation. Since the pairing in the 1PDM is between alpha and beta spins, the UHF approach only describes opposite-spin correlation, whereas through inter-pair coupling the GHF approach possesses the flexibility to describe same-spin correlation as well.

\subsection{Projected Hartree-Fock}

The primary drawback of broken symmetry wave functions lies in the difficulty of recapturing  quantum numbers associated with the broken symmetries. Many attempts have been made to address this problem, \cite{Goddard:1968dl,Mayer:1971je,Mayer:1973wx,Mayer:1973eq,Rosenberg:1975dv,Klimo:1978cg} originating with the work of L\"{o}wdin.\cite{Lowdin:1955zz} Recently a significant breakthrough was made in determining, self-consistently, symmetry adapted wave functions from a single broken-symmetry (deformed) determinant. \cite{Scuseria:2011ig} The resulting wave function possesses multi-reference character calculated with \textit{mean-field} computational cost. This projected quasi-particle theory (PQT) relies upon the simple idea of deliberately breaking and then self-consistently restoring the symmetries of the Hamiltonian with the use of projection operators. A computationally efficient formulation of projected Hartree-Fock theory based upon PQT recently appeared in the literature. \cite{JimenezHoyos:2012gx} In that work, breaking and restoring discrete symmetries such as point group and complex-conjugation symmetries were also addressed. In this work, we focus on spin symmetries as they are most familiar to quantum chemists, though the method employs an approach analogous to that of restoration of angular momentum in the context of nuclear physics. \cite{Sheikh:2000uo} 

Let us begin by taking a moment to review the two types of wave functions we will project. The collinear UHF wave function $|UHF\rangle$ minimizes the energy functional:
\begin{equation}\label{eq:Euhf}
	E = \frac{\langle \Psi | \hat{H} | \Psi \rangle} {\langle \Psi  | \Psi \rangle},
\end{equation}
sacrifices spatial and spin symmetry, and retains $\hat{S}_{z}$ as a good quantum number. The non-collinear GHF wave function minimizes the same functional with the additional sacrifice of the $\hat{S}_{z}$ quantum number. We will denote projections of these two wave functions as $|SUHF\rangle$ and $|SGHF\rangle$, respectively. We cannot emphasize enough that these projected wave functions are not single Slater determinants. Consequently, the density matrices associated with these \textit{multi-reference} wave functions are \textit{not} idempotent, though they do preserve the symmetries of the Hamiltonian.

To determine our spin-projected wave functions, we write the energy of a projected, deformed determinant $|\Phi\rangle$ as:
\begin{equation}\label{eq:Ephf}
	E = \frac{\langle \Phi | \hat{P}^{\dagger} \hat{H} \hat{P} | \Phi \rangle}{\langle \Phi | \hat{P} | \Phi \rangle},
\end{equation}
where $\hat{H}$ is the Hamiltonian, and $\hat{P}$ is the (Hermitian and idempotent) projection operator that restores the symmetries broken in the deformed determinant $|\Phi\rangle$. Formally, for continuous symmetries such as $\hat{S}^{2}$ and $\hat{S}_{z}$, the projection operator takes the form of an integral. In practice, we discretize this integral over a grid of modest size. 

It can then be shown that the energy expression \eqref{eq:Ephf} can be written as a functional of the one-particle density matrix, $\bm{\rho}_{\Phi}$, associated with the underlying deformed determinant $|\Phi\rangle$. One then minimizes the resulting functional with respect to the deformed orbitals. The resulting PHF equations are qualitatively similar to that of Hartree-Fock and the method retains \textit{mean-field} computation cost. Further, as with Hartree-Fock, any symmetry present in the initial guess for the deformed determinant will be preserved throughout the optimization procedure.\cite{JimenezHoyos:2011ia} 

We pause for a moment to remark that the implementation described in ref. (\citenum{JimenezHoyos:2011ia}) does not strictly restore $\hat{S}^{2}$, as this would require a two-body operator. Instead, it restores rotational invariance of the deformed determinant in spin-space. For collinear determinants, we rotate the determinant so that it lies along the z-direction, and then make it rotationally invariant using a projection operator of the form:
\begin{equation}
    \hat{P} = \frac{2s + 1}{2} \int_{0}^{\pi} d\beta \sin(\beta) d^{s}_{mm}(\beta)e^{i\beta \hat{S}_{y}},
\end{equation}
where $ d^{s}_{mm}(\beta)=\langle s;m|e^{i\beta \hat{S}_{y}}|s;m\rangle$. This leaves a particular direction as ``special'' in that the projected wave function (and its energy) depends on the choice of the quantum number $m = \langle\hat{S}_{z}\rangle$ of the collinear determinant. Projecting a non-collinear determinant requires integration over three Euler angles and eliminates the dependence on the quantum numbers of the deformed state.

Interestingly, the projected wave function, though multi-reference in nature, may be characterized by a single, deformed determinant we denote as $|\Phi\rangle$ or, equivalently, the associated one-particle density matrix $\bm{\rho}_{\Phi}$. We want to stress that $|\Phi\rangle \ne |UHF\rangle$ or $|GHF\rangle$. The determinants $|UHF\rangle$ or $|GHF\rangle$ optimize the energy functional of Equation \eqref{eq:Euhf} (breaking either $\hat{S}^2$, or $\hat{S}^2$ and $\hat{S}_z$, respectively), whereas the determinant $|\Phi\rangle$ optimizes the energy functional of Equation \eqref{eq:Ephf}. The determinant $|\Phi\rangle$ can be of either the UHF or GHF type, depending on what symmetries are broken. 

Finally, we remark that though the SUHF wave function and the spin projected extended Hartree-Fock (EHF) method are in fact equivalent, we obtain our results without working in the corresponding orbital basis (as suggested by Mayer and L\"{o}wdin) resulting in significantly less computational overhead.\cite{JimenezHoyos:2012gx} In addition, previous examples of PHF in the literature have all been collinear, i.e. restoring only $\hat{S}^{s}$, rather than both $\hat{S}^{s}$ and $\hat{S}_{z}$, i.e. previous work did not consider SGHF wave functions. As Pulay pointed out, \cite{Pulay:1988ic} when several orbitals are strongly correlated, the EHF (equivalently the SUHF) wave function doesn't have the flexibility to describe all of the static correlation in the system. The SGHF wave function provides a way around this: by breaking and restoring $\hat{S}_z$ as well as $\hat{S}^2$, it captures inter-pair correlation which the SUHF wave function cannot.

\section{Hydrogen Networks\label{sec:Results}}

\subsection{Computational Details}

We utilize an implementation of the PHF equations in the development version of the {\sc{gaussian}} suite of programs.\cite{gdv-H1} A minimal (STO-3G) basis set was deemed sufficient for the purpose of describing the qualitative features of strong correlation associated with near degeneracies in the orbital spectrum. The initial guess for deformed SUHF determinants are broken symmetry Hartree-Fock states. The initial guess for deformed SGHF calculations were constructed using an SUHF determinant and mixing the alpha and beta orbitals nearest the Fermi level with a small mixing angle.

\subsection{H$_{4}$ and H$_{6}$ Rings}

\begin{figure}[ht]
	\centering
	\includegraphics[width=2.9in]{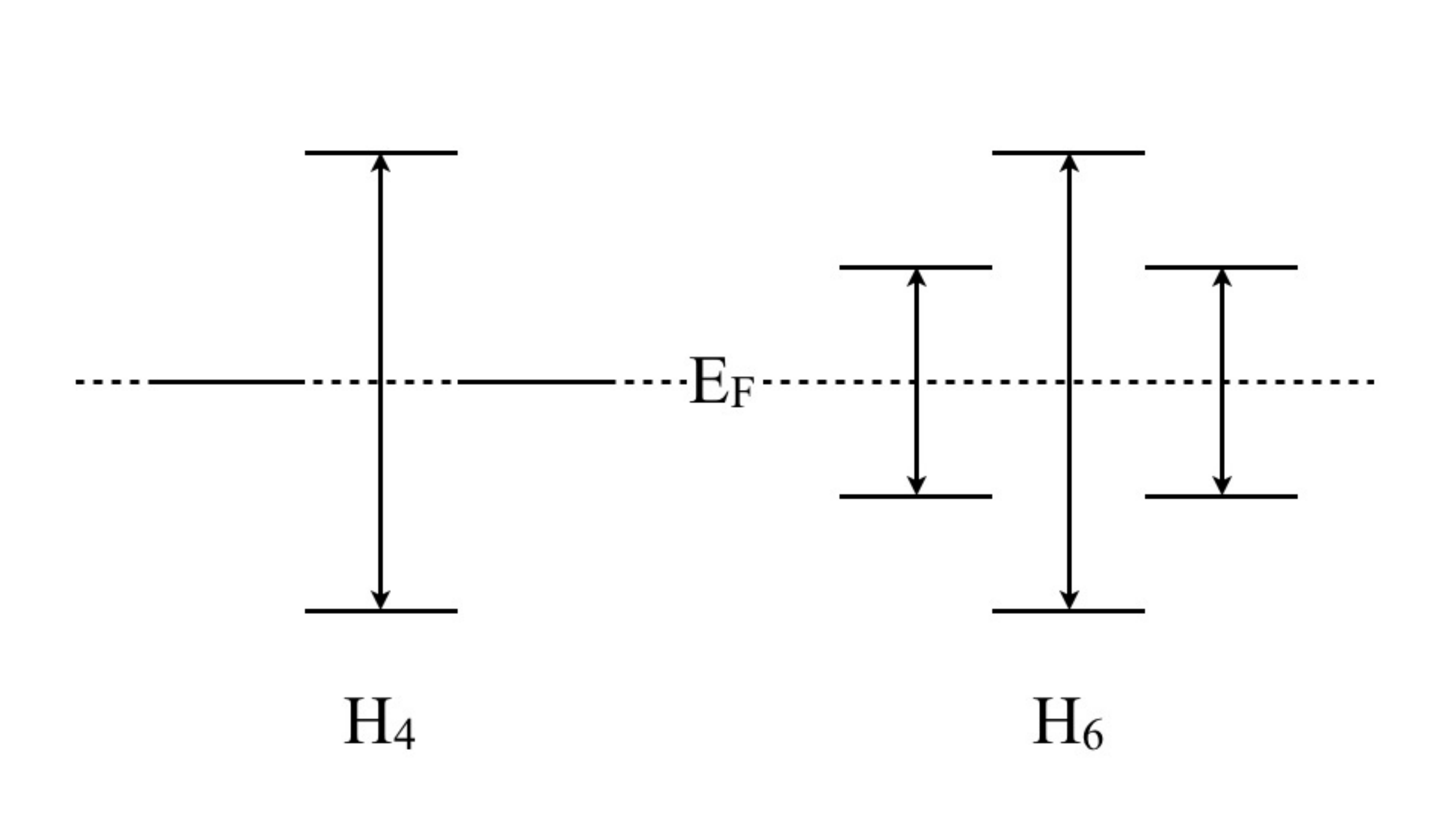}
	\caption[]{\label{fig:orbital-schematic}
	A schematic depiction of the one-electron molecular orbital energy spectrum for the H$_{4}$ (left) and H$_{6}$ (right) rings, with L\"{o}wdin's pairing scheme indicated by arrows. The effective Fermi energy (E$_{F}$) lies in the degenerate orbitals for H$_{4}$ and in the gap in H$_{6}$.
	}
\end{figure}

We illustrate the evolution of pair wave functions as a function of nearest-neighbor distance ($R_{n-n}$), as well as investigate the errors associated with different types of spin projection (SUHF and SGHF), by considering the symmetric stretching of rings of four and six hydrogen atoms. There is a topological difference between the orbital spectra of the four and six membered rings, and in general between 4N and 4N+2 membered rings, indicated in Figure \ref{fig:orbital-schematic}. In general, 4N membered (anti-resonant) rings have an effective Fermi level which lies within two half-filled degenerate orbitals, whereas the effective Fermi level for the 4N+2 membered (resonant) rings lies between a pair of fully occupied degenerate orbitals and a pair of unoccupied degenerate orbitals. One expects that the ground state for both systems should be a singlet, though in the case of anti-resonant rings, an open-shell singlet.

\begin{figure}[ht]
	\subfloat[][]{
	\centering
	\includegraphics[width=2.9in] {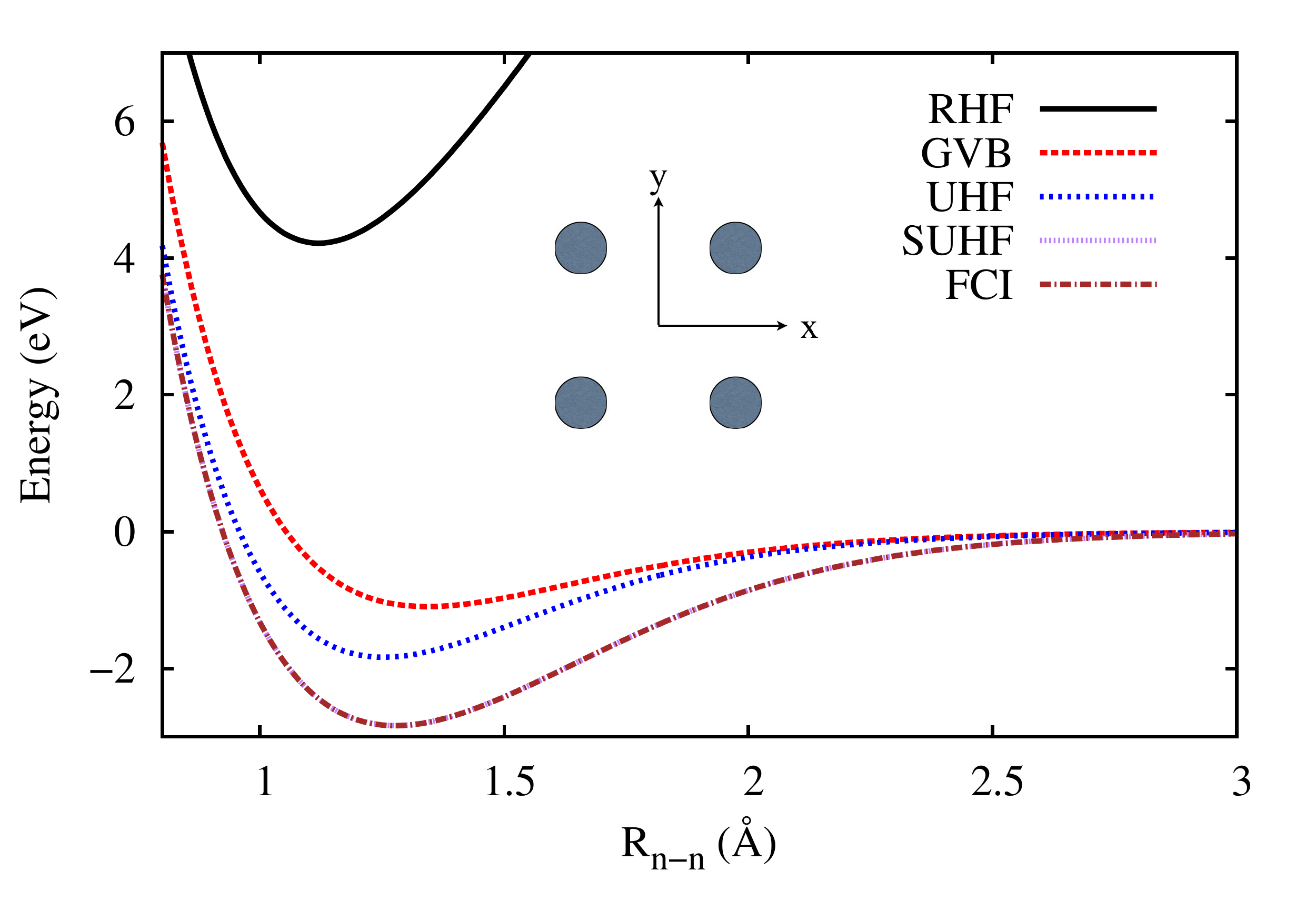}
	\label{fig:H4-s-en}
	}\\
	\subfloat[][]{
	\centering
	\includegraphics[width=2.9in] {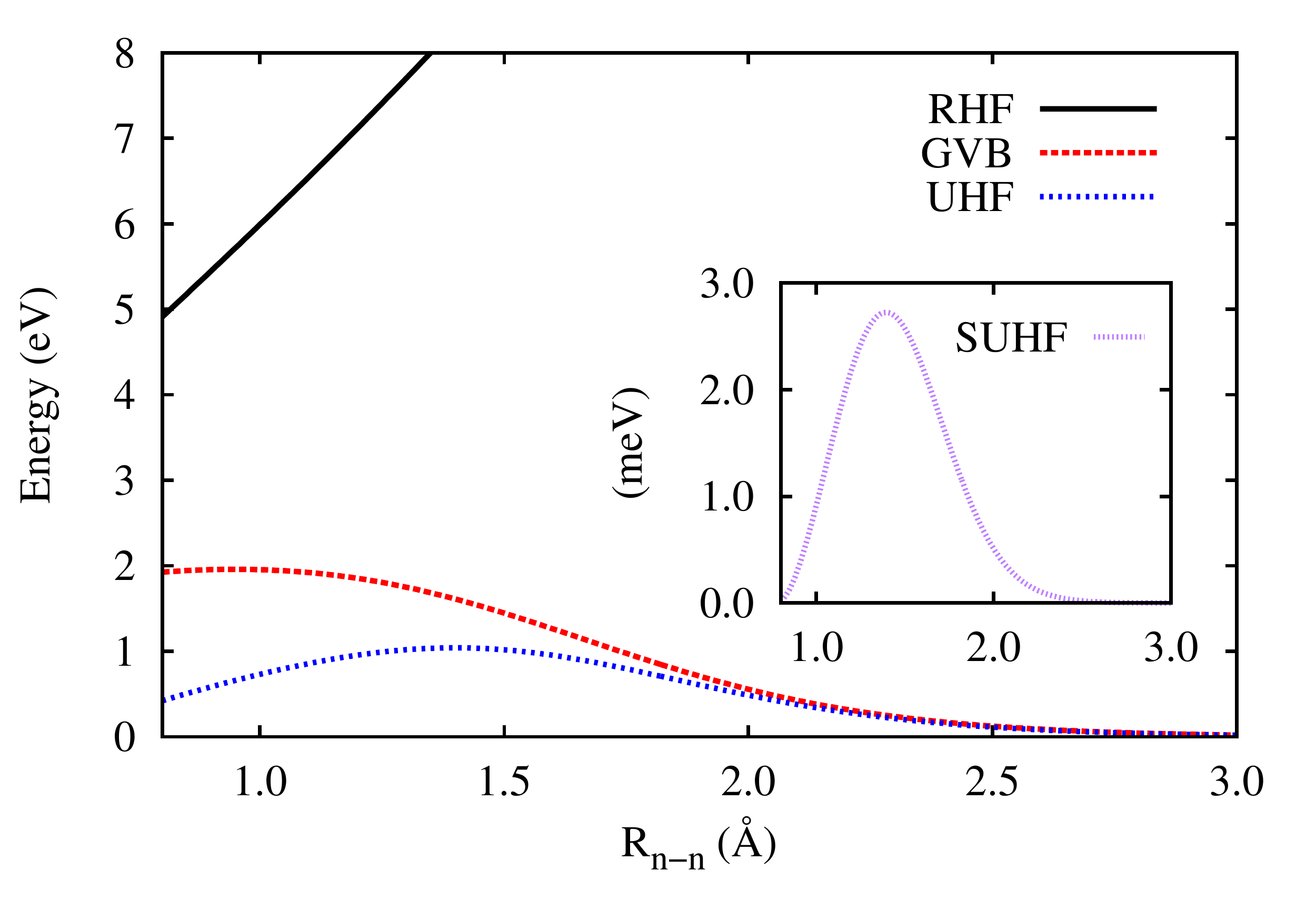}
	\label{fig:H4-s-fci}
	}
	\caption[]{
	 \subref{fig:H4-s-en} Dissociation curves for different singlet wave functions for the ground state of the H$_4$ square in a minimal basis. The zero of energy has been set at the energy of four Hartree-Fock hydrogen atoms. (inset) Arrangement of the atoms with respect to x and y symmetry axes. \subref{fig:H4-s-fci} Energy difference from FCI of several singlet wave functions for H$_{4}$. The SUHF energy remains within 3 meV of the exact solution at all distances. 
    }
\end{figure}

The ground state of H$_{4}$ adopts a $1a_{g}^{2}(e_{x},e_{y})^2$ configuration, with $x$ and $y$ defined in the inset of Figure \ref{fig:H4-s-en}. The open-shell configuration gives rise to three singlet states: $^{1}A_{g} (x^{2}+y^{2})$, $^{1}B_{2g} (xy)$, and $^{1}B_{1g} (x^{2}-y^{2})$, and a triplet state: $^{3}B_{2g}(xy)$. Placing the electrons in $^1B_{1g} (x^2-y^2)$ yields the lowest energy. The reason can be understood in the following way: the $^{1}B_{1g}$ state takes the form of an open-shell singlet $e_{x+y}^1, e_{x-y}^1$, where one electron resides on each of the atoms lying on one diagonal of the square (x+y), and the other along its complement (x-y). This segregates the alpha and beta spins onto separate sublattices allowing them to avoid double occupancy and the consequent on-site repulsion. The electrons in the lowest ($1a_g$) orbital then correlate by mixing with its anti-bonding counterpart ($1b_g$) as shown in Figure \ref{fig:orbital-schematic}.

We present the $^1B_{1g}$ dissociation curves for several approximations in Figure \ref{fig:H4-s-en} as a function of nearest-neighbor distance ($R_{n-n}$). While the RHF approximation dissociates improperly, both the UHF and GVB approximations succeed. Figure \ref{fig:H4-s-fci} shows the energy difference from the exact (FCI) result and one can see that, in the intermediate coupling regime ($R_{n-n} < 2.0 $\AA), UHF has a significantly smaller error than GVB. This reflects the advantage of allowing orbital spin contamination in the UHF approach. Neither of these wave functions are equal to the FCI for this system, and the error in the absolute energy is on the order of 1-2 eV near equilibrium.

In the intermediate recoupling regime where both UHF and GVB still have significant error, the projected wave functions offer significant improvement in the energy, in addition to restoring the correct quantum numbers. This improvement results from the multi-reference nature of the projected wave function. The error in the SUHF wave function reduces to the order of meV, and the SGHF wave function is identical to the FCI for the minimal basis. Recall that for \ce{H2} both the SUHF and the GVB-PP approximations are identical to the exact wave function for the minimal basis set. While they continue to reflect the appropriate intra-pair correlation in the four electron system, they do not treat the inter-pair correlation completely, a problem ultimately associated with the lack of size-consistency in these methods.

To understand these results, recall that there exist two distinct singlet spin states for any four-particle fermionic system:\cite{Amos:1975ky}
\begin{subequations}\label{eq:4part_singlet}
\begin{align} 
	|1S\rangle &= \szero \szero\\
    \begin{split}
	|2S\rangle &= \left(\frac{1}{\sqrt{3}}\right)\tup\tdown \\
     &+ \left(\frac{1}{\sqrt{3}}\right)\tdown\tup \\ &- \left(\frac{1}{\sqrt{3}}\right)\tzero\tzero.
     \end{split}
\end{align}
\end{subequations}
Here, $|1S\rangle$ describes the product of two singlet coupled pairs as in the GVB-PP wave function, while $|2S\rangle$ describes a linear combination of products of triplet pairs coupling to form an overall singlet state. The dominant contribution to the ground state comes from $|1S\rangle$, though the exact wave function does contain a small contribution from $|2S\rangle$, particularly near the equilibrium bond length. The GVB-PP wave function successfully describes all of the contribution to the total energy associated with the $|1S\rangle$ spin state. The UHF singlet obtains a better total energy by breaking spatial symmetry and introducting spin contamination. 

For the spin projected states, the SUHF wave function captures correlation from determinants that couple to the $|1S\rangle$ spinor. Though both GVB-PP and SUHF wave functions are multi-reference and both couple to $|1S\rangle$, SUHF captures significantly more correlation because it contains more determinants than the GVB-PP. The first two terms of $|2S\rangle$ describe \textit{same-spin} correlation, i.e. contributions from the product of $m_{s} = \pm 1$ two-electron triplets. There are no same-spin correlations present in a deformed $m_{s} = 0$ singlet determinant; consequently, the SUHF state cannot describe correlation from determinants that couple to $|2S\rangle$, resulting in the energy difference from FCI seen in Figure \ref{fig:H4-s-fci}. Breaking $\hat{S}_{z}$ as well as $\hat{S}^2$, allows the SGHF wave function to further correlate the two pairs by capturing this same-spin correlation. By \textit{deliberately} breaking and restoring this symmetry, the SGHF wave function recovers the full inter-pair correlation energy, i.e. the FCI result. 

\begin{figure}[ht]
	\subfloat[][]{
		\centering
		\includegraphics[width=2.9in]  {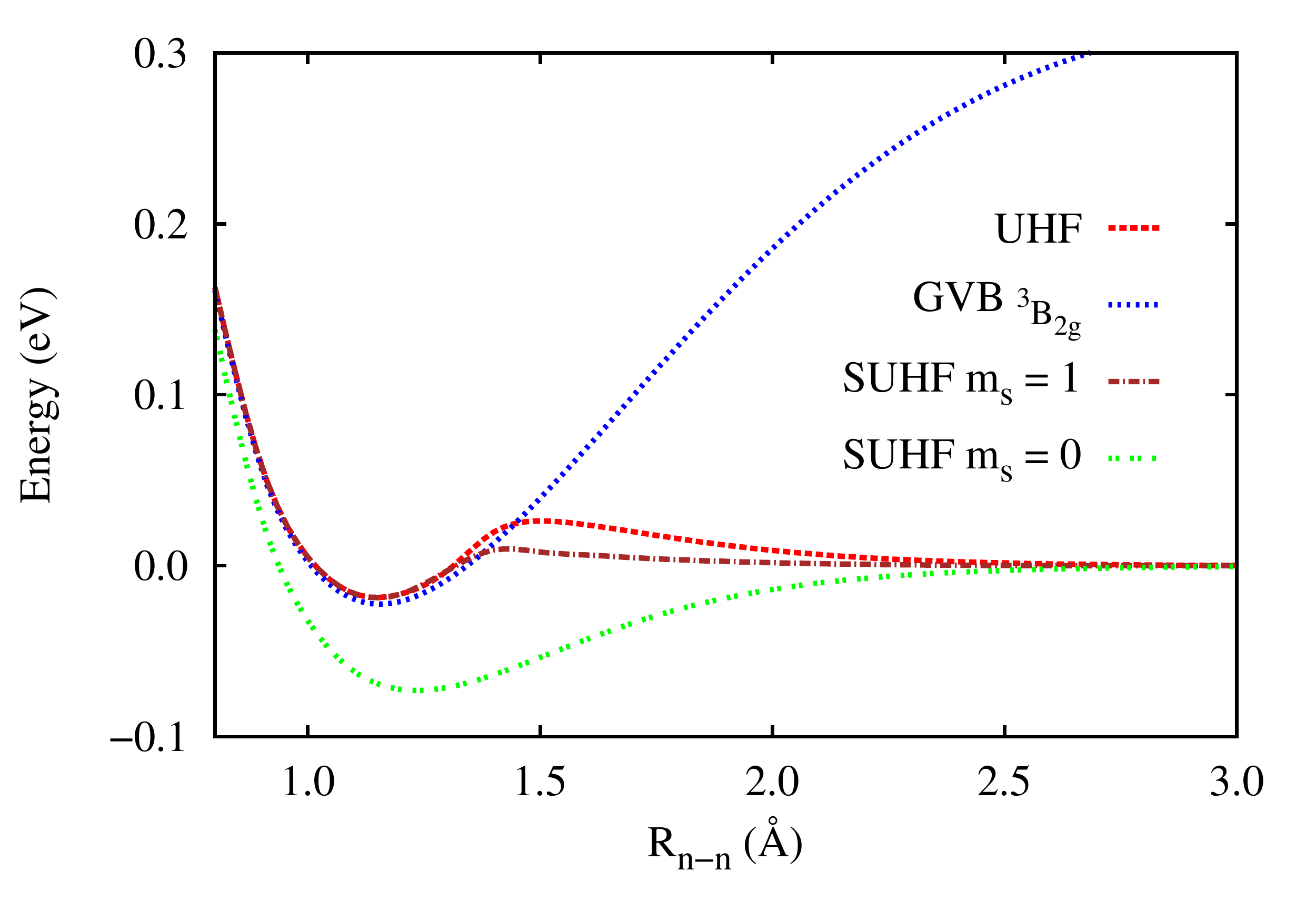}
		\label{fig:H4-t-en}
	}\\
	\subfloat[][]{
		\centering
		\includegraphics[width=2.9in] {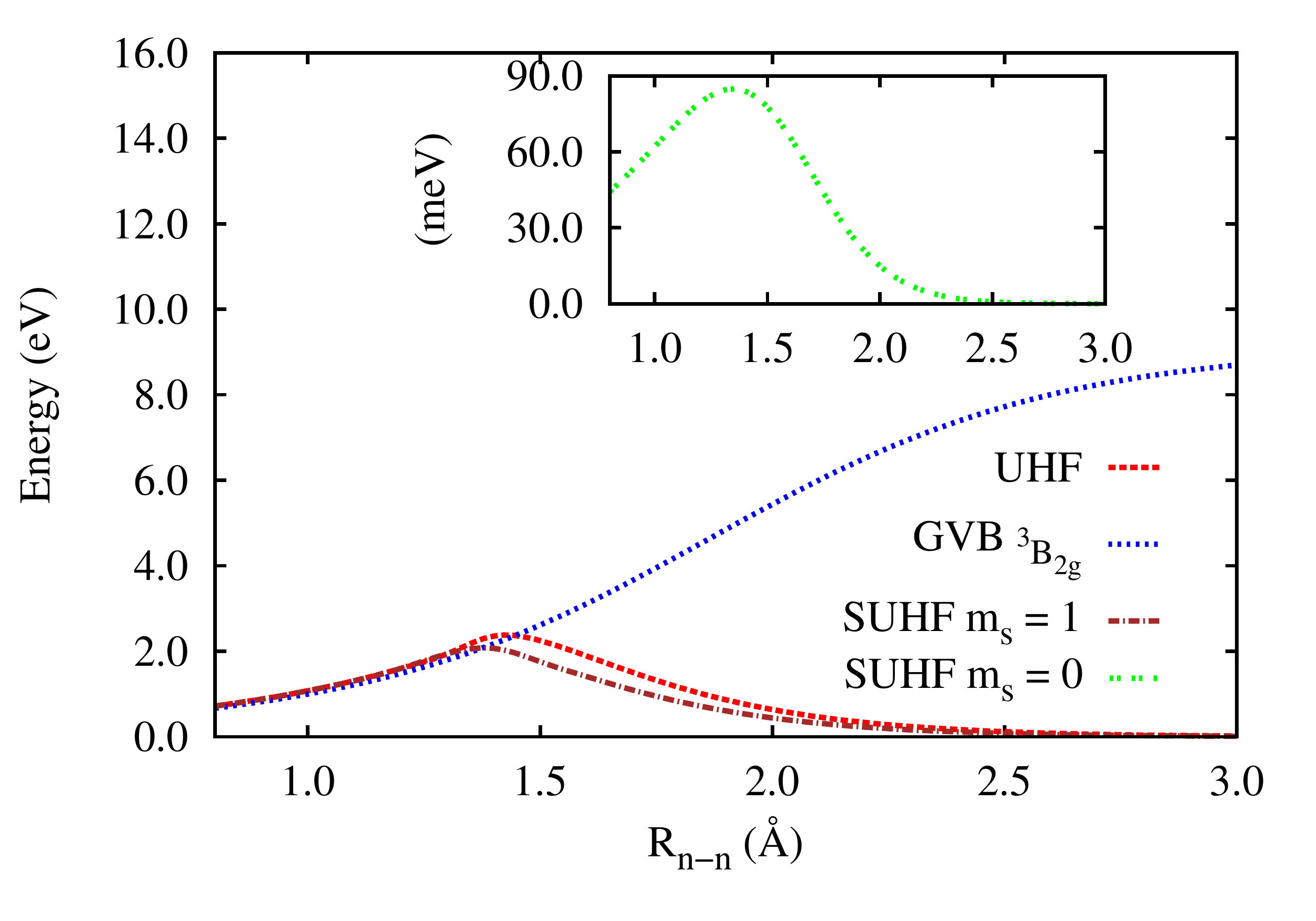}
		\label{fig:H4-t-fci}
	}
	\caption[]{\label{fig:H4t}  \subref{fig:H4-t-en} Dissociation curves for an H$_4$ square for different triplet wave functions in a minimal basis. The zero of energy has been set at the energy of four Hartree-Fock hydrogen atoms. \subref{fig:H4-t-fci} Energy difference from FCI for the same wave functions.}
\end{figure}

There is another point we wish to make with this simple model exemplified in the triplet potential curves depicted in Figure \ref{fig:H4t}. The GVB-PP $^3B_{2g}(xy)$ triplet state arising from the ground state configuration has the wrong symmetry to dissociate properly as evident in Figure \ref{fig:H4-t-en}. Another triplet state, based on the excited electronic configuration $1a_g^1(e_x,e_y)^3$ (not shown) does, but the dissociation curve for this state is purely repulsive and not appropriate at equilibrium. The UHF solution, by breaking spatial symmetry, yields a qualitatively correct curve at all distances, though it predicts a barrier to formation of the bound state.  

Turning now to the projected wave functions, we see that, as in the singlet case, the SGHF triplet wave function is equal to FCI for this minimal basis. However, we find two distinct SUHF triplet wave functions with very different energies; one arises from projecting a deformed determinant with $m_{s} = 0$ and a second arises from projecting a deformed determinant with $m_{s} = \pm 1$. As we briefly mentioned above, these two states differdue to the fact that we simply assure rotational invariance of the projected state, without respect for a particular axis in spin space. When we restore both rotational invariance as well as $\hat{S}_{z}$ (i.e. SGHF) we don't have this problem. 

Using this simple four-electron system, we can understand why this occurs. The PHF equations preserve the symmetries of the initial guess throughout the SCF procedure. The symmetries of the two deformed determinants are different and thus capture different types of correlation. Recall that there exist three distinct four-particle triplet spinors:\cite{Amos:1975ky}
\begin{subequations}
\begin{align}
	|1T\rangle &= \szero\tzero\\
	|2T\rangle &= \tzero\szero\\
	|3T\rangle &= \frac{1}{\sqrt{2}}\left(\tup\tdown - \tdown\tup\right)
\end{align}
\end{subequations}
We obtain the SUHF $m_{s} = 1$ wave function by taking the UHF open-shell triplet state as an initial guess. For distances smaller than the triplet CF point, the deformed determinant has a doubly occupied $1a_{g}$ orbital and the open-shell electrons occupy the $e_{x}$ and $e_{y}$ orbitals, delocalizing the un-paired spins across all the atoms. Beyond the triplet CF point, the deformed determinant breaks symmetry, mixing the $e_{x}$ and $e_{y}$ orbitals as described above to localize the open-shell spins on different atoms. We therefore see that beyond the CF point, the two pairs must form the product spinor $|3T\rangle$, as the open-shell geminal's spinor takes the form of either $\tup$ or $\tdown$, and the remaining geminal couples to the spinor: $\frac{1}{\sqrt{2}}\left(\tup - \tdown\right)$. This provides only a modest improvement in the energy compared to UHF as symmetry breaking captures most of the correlation.

In contrast, we obtain the SUHF $m_{s} = 0$ wave function state by taking the UHF singlet state as an initial guess. This deformed determinant captures contributions from $|1T\rangle$ and $|2T\rangle$, and the resulting SUHF state captures determinants which couple to product geminals from the open-shell singlet and the paired triplet (and vice versa). As there are significantly more determinants in the FCI space which couple to these four-particle spinors, the maximum error drops from almost 2 eV to slightly less than 90 meV (see Figure \ref{fig:H4-t-fci}, inset).

\begin{figure}[ht]
	\centering
	\includegraphics[width=2.9in]{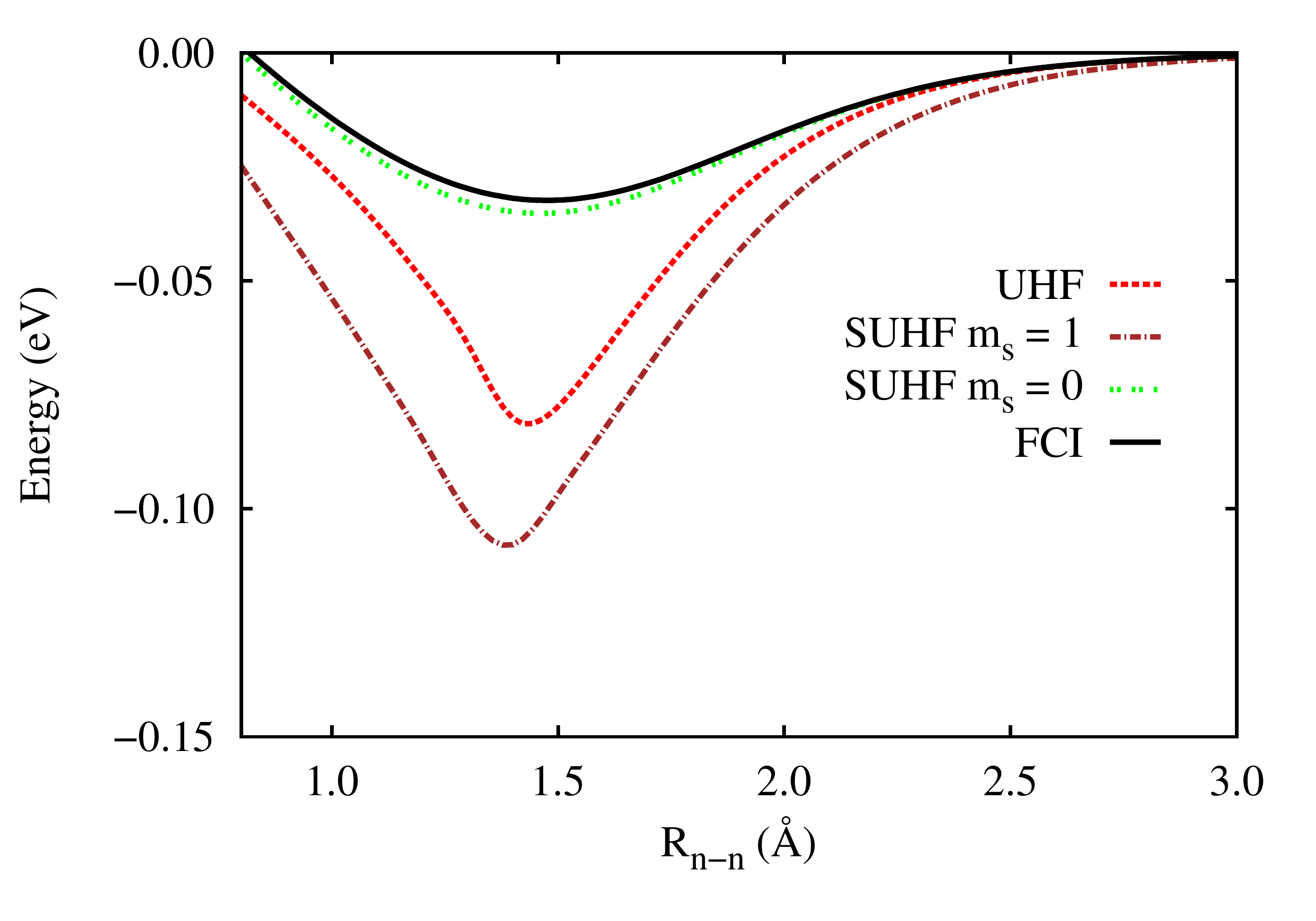}
	\caption[]{ The singlet-triplet energy difference for H$_4$. Sharp cusps in the UHF and SUHF $m_{s}=1$ states are the result of spontaneous symmetry breaking at a CF point.}
	\label{fig:h4-st-split}
\end{figure}

This has consequences for calculating magnetic properties of materials. We plot the singlet-triplet splitting for the H$_4$ system in Figure \ref{fig:h4-st-split}. At large $R_{n-n}$ this splitting relates to the magnetic coupling constants of the Heisenberg model Hamiltonian. The dissociation curves of both the UHF and the SUHF $m_{s} = 1$ state have a CF point, and consequently the singlet-triplet splitting calculated from these states possesses a discontinuity in the derivative at that point. Combined with the significant underestimation of the triplet energy, the splitting calculated using the SUHF $m_{s} = 1$ wave function is actually worse than using the UHF single determinant. In contrast, the $m_{s} = 0$ SUHF state gives both a quantitatively as well as qualitatively correct curve.

\begin{figure*}[ht]
	\centering
	\includegraphics[width=6.0in]{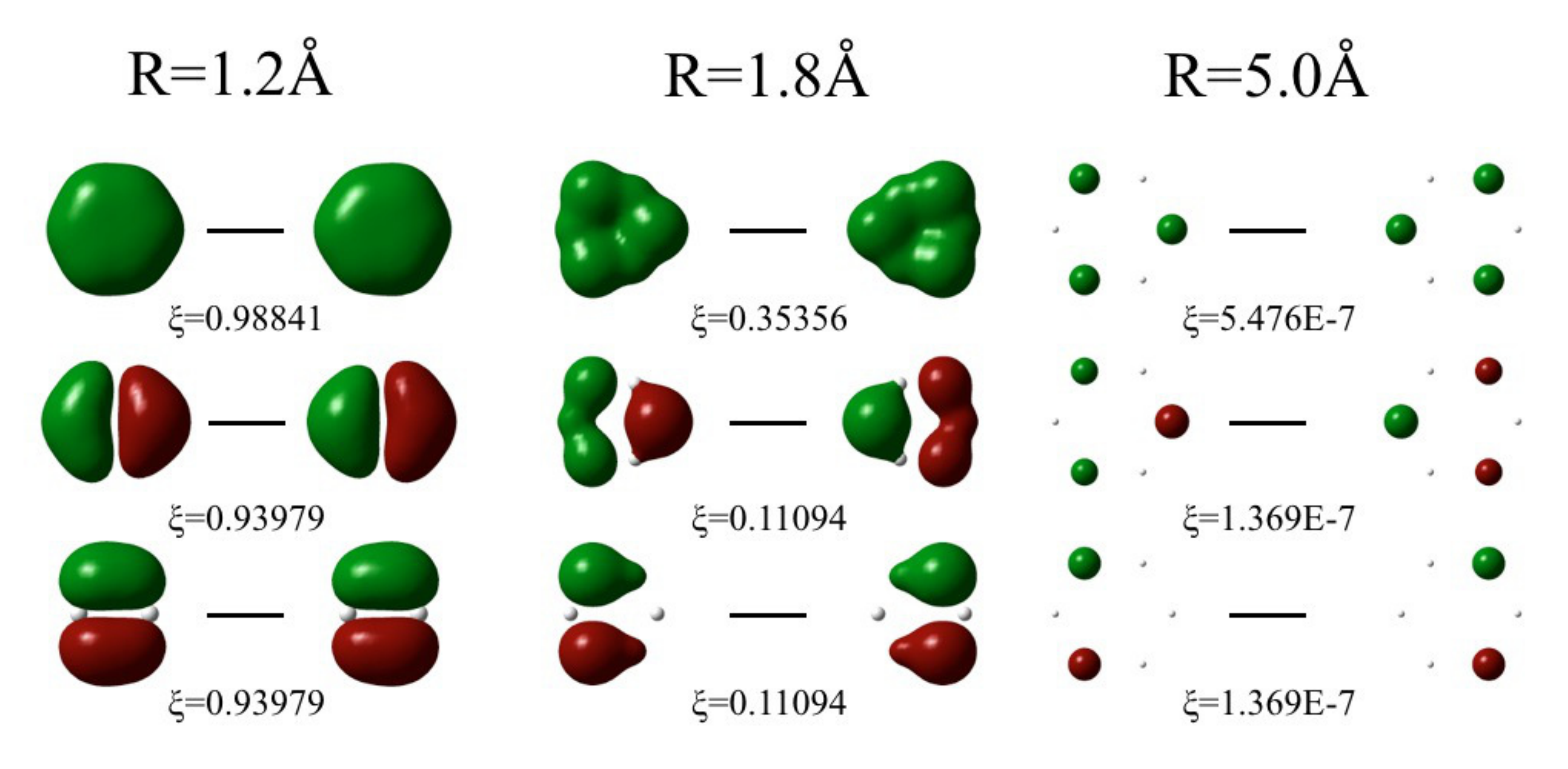}
	\caption[]{ UHF corresponding (pair) orbitals for the H$_{6}$ ring at short, intermediate and large separation.\bibnotemark[sym] We present the pairs with the $\alpha$ orbital on the left and the corresponding $\beta$ orbital on the right; $\sqrt{\xi}$ is the orbital overlap (see text).}
	\label{fig:h6-pairs}
\end{figure*}

Turning now to $H_{6}$, two additional electrons fill the degenerate $e_{i}$ orbitals, leading to a closed-shell singlet ground state. In Figure \ref{fig:h6-pairs}, we plot the UHF pair functions for the three pairs of $H_{6}$ ring at three distances: 1.2 \AA, 1.8 \AA, and 5.00 \AA, corresponding to just past the CF point, moderately stretched, and dissociation. The pairs evolve smoothly as the $\alpha$ and $\beta$ electrons segregate onto two distinct sublattices. In order to understand the manner in which the pair functions evolve, we project the pair functions back onto the symmetry adapted MOs of the system. Orbital coefficients for the alpha and beta pairs for this system \bibnote[sym]{The perfect geometric symmetry of the ring was broken slightly by adding a small ($0.01$\ \r A) displacement to each succeeding atom along the vertical z-axis. This imposed helical distortion allows us to uniquely define each of the molecular orbitals of the system, as it lifts the perfect degeneracy in the $e$ symmetries, and removes the freedom of the diagonalization routine to rotate them in an arbritrary manner.} in the symmetry adapted MO basis at a distance of 1.80 \AA \ appear in Table \ref{tab:coeff}, showing how the pairs form by mixing a bonding orbital with its anti-bonding counterpart. In the dissociation limit, the orbital coefficients approach the limiting value of $1/\sqrt{2}$. The pairing ansatz evident in the UHF wavefunction, where each orbital mixes with its corresponding anti-bonding orbital, was suggested as a way to correlate electrons subject to large Coulomb interactions by L\"{o}wdin in 1955 and goes by the name of alternant molecular orbital (AMO) theory. \cite{pauncz1967alternant,Lowdin:1955zz} For the projected wave functions, the underlying deformed determinant of the SUHF state possesses pairing scheme.

\begin{table*}[ht]
	\begin{tabular}{|c|c|c|c|c|c|c|}
	\hline\hline
	& \multicolumn{3}{c|}{Alpha} &\multicolumn{3}{c|}{Beta}\\
	\hline
	MO & Pair 1 & Pair 2 & Pair 3 & Pair 1 & Pair 2 & Pair 3 \\
	\hline
	$A_{1g}$ & -0.893  & 0     &  0      & -0.893  & 0       &  0      \\
	$E_{1u}$ &  0      & 0.816 &  0      &  0      & \ 0.816 &  0      \\  
	$E_{1u}$ &  0      & 0     & -0.816  &  0      & 0       & -0.816  \\
	$E_{2g}$ &  0      & 0     & \ 0.577 &  0      & 0       & -0.577  \\
	$E_{2g}$ &  0      & 0.577 &  0      &  0      & -0.577  &  0      \\
	$B_{1u}$ & \ 0.450 & 0     &  0      & -0.450 & 0       &  0      \\
	\hline
	\end{tabular}
\caption[]{
	\label{tab:coeff}
	Corresponding pair coefficients in the symmetry adapted basis\bibnotemark[sym] for H$_{6}$ at intermediate (1.80 \AA) bond length. Orbitals pair as shown in Figure \ref{fig:orbital-schematic}.
	}
\end{table*}
To summarize, we believe these approaches capture the most important static correlation effects. With the SUHF wave function, we capture all the intra-pair correlation effects, and the SGHF approach captures additional pair-pair correlation. Even so, higher order many-body effects are not described properly by these methods, and those terms grow in importance as the system size increases, leading to a lack of size extensivity.\cite{JimenezHoyos:2012gx}

\subsection{Larger Rings}
\label{subsec:rings}

We now consider larger 1D rings of hydrogen as a function of nearest-neighbor bond length. We present results for both 16 and 18 atom rings; chosen because they are complicated enough that FCI calculations become quite cumbersome, yet the system remains small enough that the PHF approach can provide a significant improvement over UHF, i.e. we do not approach the thermodynamic ($N_e \rightarrow \infty$) limit. Our results simply extend those already seen in the four and six atom rings, and in what follows we focus on the resonant system: \ce{H18}. 

\begin{figure}[ht]
	\subfloat[][]{
		\centering
		\includegraphics[width=2.9in] {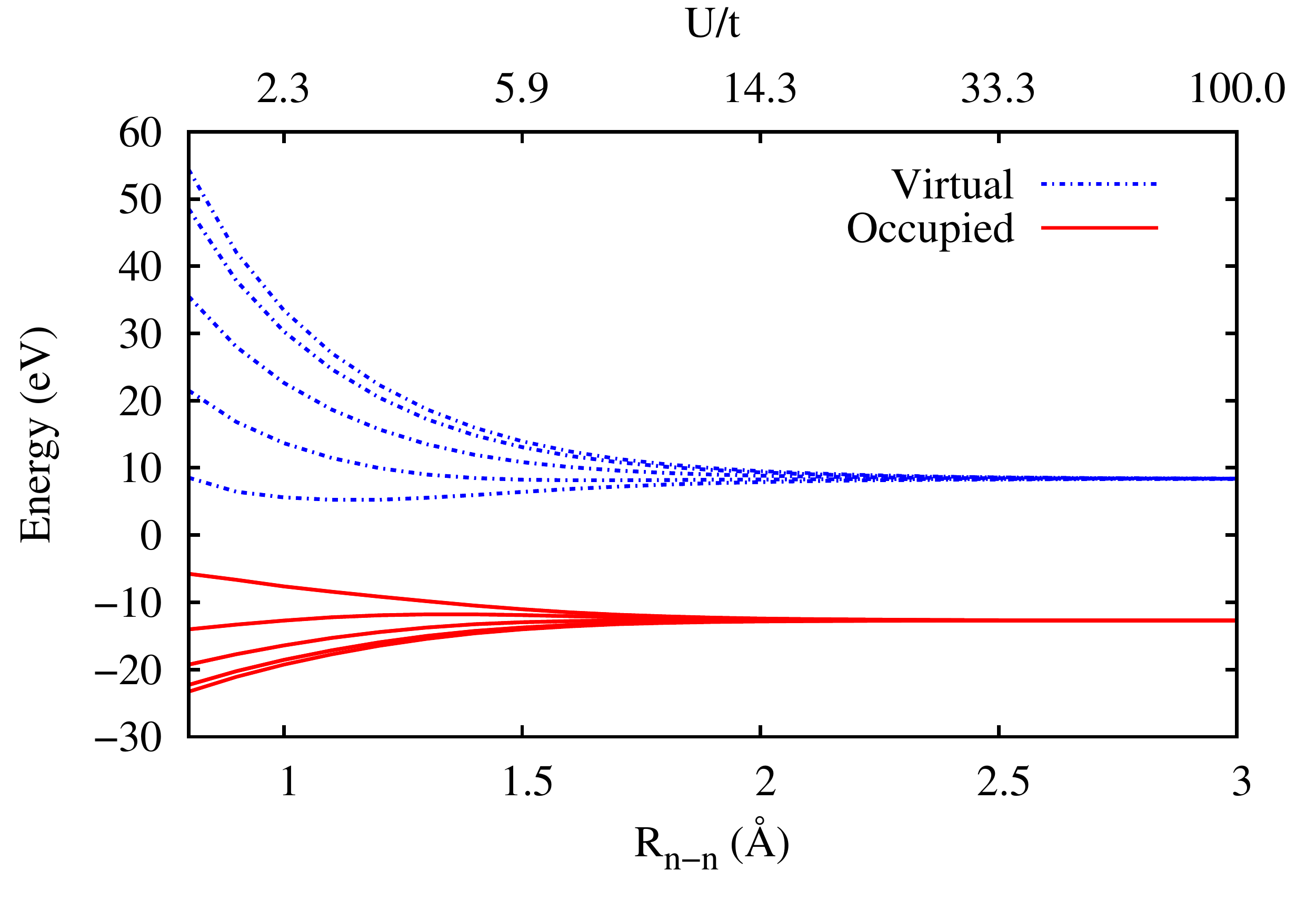}
		\label{fig:H18-spectrum}
	}\\
	\subfloat[][]{
		\centering
		\includegraphics[width=2.9in]{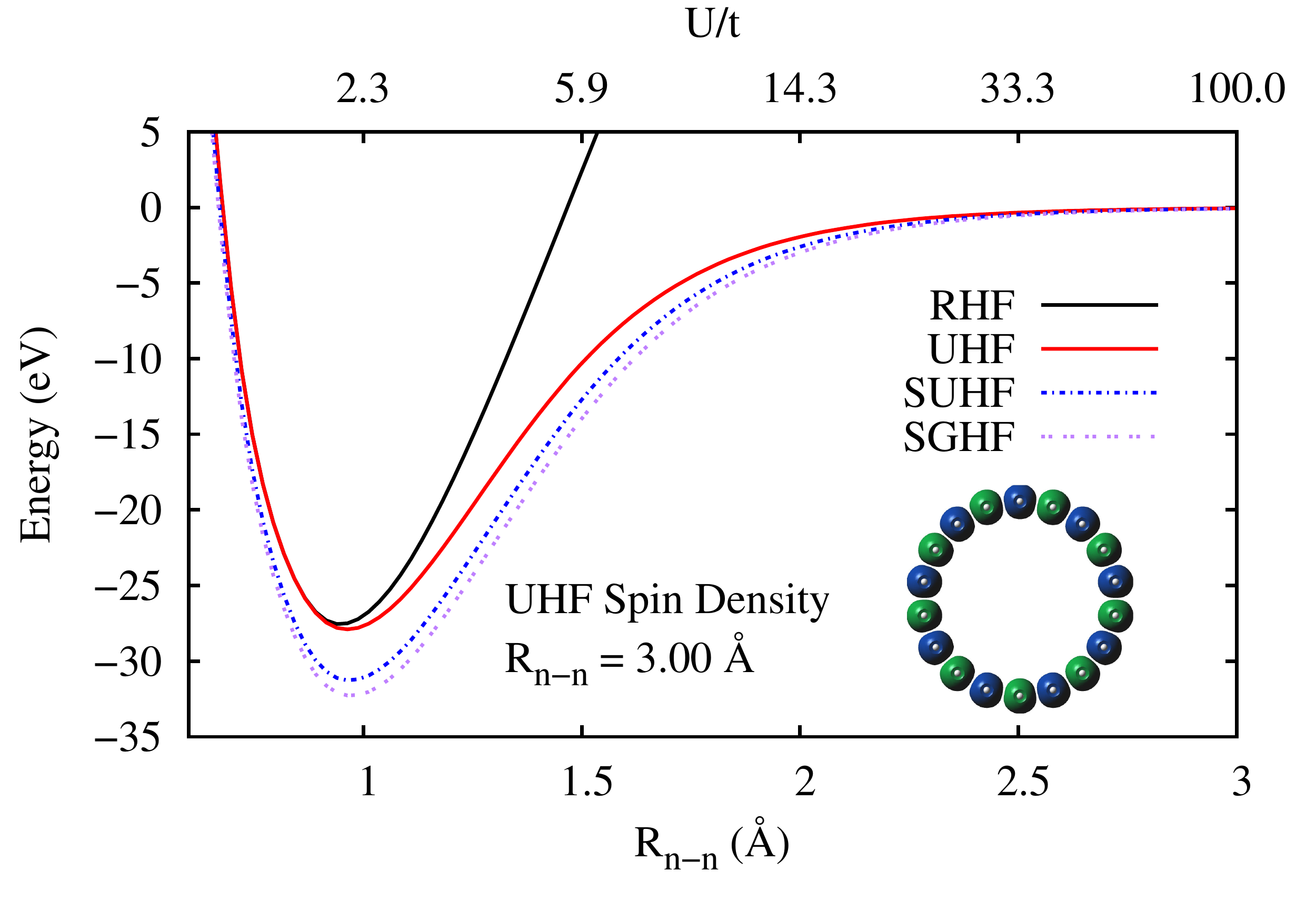}
		\label{fig:h18-dis}
	}
	\caption[]{
	\label{fig:H18-energy}
	 \subref{fig:H18-spectrum} The UHF orbital energy spectrum for the \ce{H18} ring as a function of R$_{n-n}$ and U/t. As the bonds are stretched, the occupied and virtual MOs each become energetically degenerate, forming upper (unoccupied) and lower (occupied) Hubbard  ``bands.'' \subref{fig:h18-dis} Dissociation curves and (inset) UHF spin density for the \ce{H18} ring. The UHF wave function localizes the electrons anti-ferromagnetically at large $R_{n-n}$.}
\end{figure}

Figure \ref{fig:H18-spectrum} shows the UHF one-electron molecular orbital energy spectrum for the \ce{H18} ring. In the small $R_{n-n}$ regime, all the orbitals except for the lowest occupied and highest virtual orbitals remain doubly degenerate. As the bonds stretch, the orbital spectrum compresses as the occupied and virutal MOs become energetically degenerate. At dissociation, the orbital eigenvalues collapse into two sets of degenerate levels, an occupied ``lower band'' and the unoccupied ``upper band'', each containing nine spatial orbitals. The effective Fermi energy sits in the gap between these two ``bands.''

Although our calculation uses the exact Hamiltonian, we map our results as a function of distance to a Hubbard model in an attempt to make contact with the substantial body of work using that approximation. To this end, we approximate the onsite interaction parameter $U$ as the UHF ``band gap'' in the limit $R_{n-n} \rightarrow \infty$. In the case of half-filling, this energy difference corresponds to the energy needed to move an electron from one site to another. This ``gap'' has a value of $\sim$ 21 eV for both the \ce{H18} and \ce{H16} rings, not an unreasonably large value for $U$ in this minimal basis set. 

We next extract the effective near-neighbor hopping parameter $t$ from the band width of the one-electron spectrum. The effective value we seek should be renormalized by all the terms in the exact Hamiltonian except for the on-site $U$ term. To this end, at each bond length we compute the UHF eigenvalue spectrum for the high-spin, ferromagnetic state. This singly occupies each orbital, thereby eliminating the onsite Hubbard $U$ interaction, and the effective hopping parameter $t$ is then given by: 
\begin{equation}
	t \approx \frac{1}{2z} \left(\epsilon_{N} - \epsilon_{0}\right),
\end{equation}
where $\epsilon_{N}$ - $\epsilon_{0}$ is the splitting between the highest and lowest occupied MOS of the ferromagnetic state, and $z$ is the number of nearest neighbors.

The onset of strong correlations corresponds to the regime where the bandwidth and the on-site repulsion become comparable. For the 1D ring, this occurs at a ratio of $U/t \sim 4$, or $R_{n-n} \sim 1.25$\ \AA, shortly past the CF point.  As $R_{n-n} \rightarrow \infty$ the effective hopping parameter tends toward zero, reflecting an effectively infinite on-site interaction.  This transition regime corresponds to distances where the SUHF energy differs most greatly from the exact result for H$_{4}$ (the inset of Figure \ref{fig:H4-s-fci}) suggesting the importance of inter-pair correlation.

In the dissociation curves for the \ce{H18} ring shown in Figure \ref{fig:h18-dis}, the UHF wave function goes to the correct limit by allowing the alpha and beta spins to localize on different sublattices (See Fig \ref{fig:h18-dis}, inset). Notice that the CF point is quite near equilibrium, i.e. the UHF wave function starts to correlate through localization in this region. One might therefore suspect that strong correlations contribute significantly at these distances. In fact, the SUHF projected wave function captures $\sim$ 3.5 eV of energy from intra-pair correlation, and the SGHF wave function captures an additional $\sim$ 1.4 eV from inter-pair correlation at the equilibrium near-neighbor distance.

\begin{figure}[ht]
	\centering
	\includegraphics[width=2.9in] {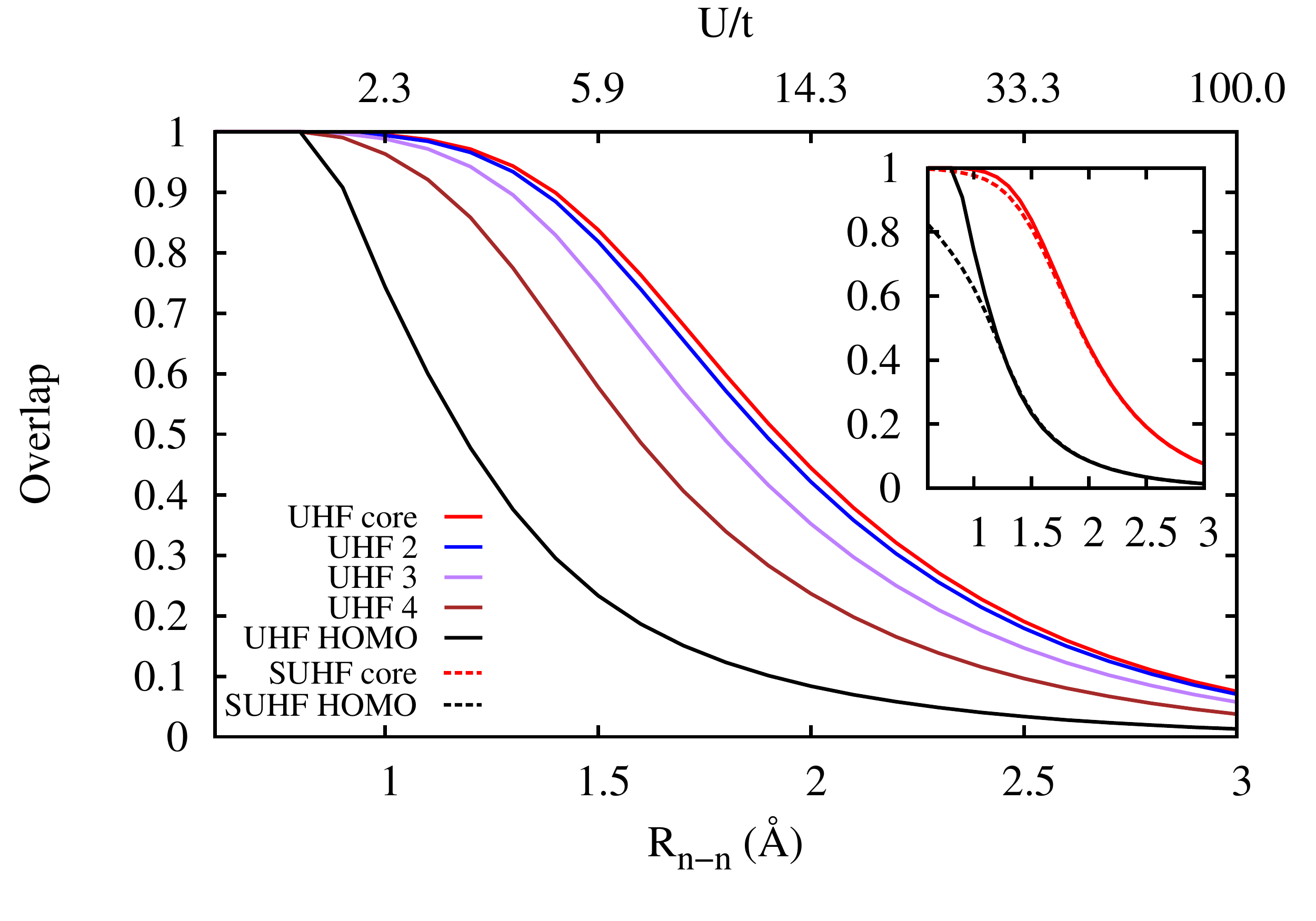}
    \caption[]{
    \label{fig:h18-uhf-over}
	 Overlap of the singlet UHF corresponding orbitals for the \ce{H18} ring (see text). All orbitals except for the core are doubly degenerate. The UHF overlap becomes identically 1 for all orbitals at the CF point. (inset) Overlap of the $\alpha$ and $\beta$ HOMO and core corresponding orbitals of the UHF and deformed SUHF determinants of the same system. 
	}
\end{figure}

The corresponding pairs in \ce{H16} and \ce{H18} are similar to those discussed for H$_4$ and H$_6$, respectively.  At small distances, the orbital overlap of each pair remains identically one (see Figure \ref{fig:h18-uhf-over}). This follows from the fact that inside the CF point, the UHF wavefunction has not yet broken symmetry and each spin-orbital ($\alpha$ and $\beta$) remain spatially identical.  The electrons nearest the Fermi level localize first as the ring stretches, and at dissociation the overlap of all the orbitals goes to zero, reflecting compete localization of the spins onto separate sublattices.

\begin{figure*}[ht]
	\subfloat[][core pair at 1 \AA]{
		\centering
		\includegraphics[width=2.9in]{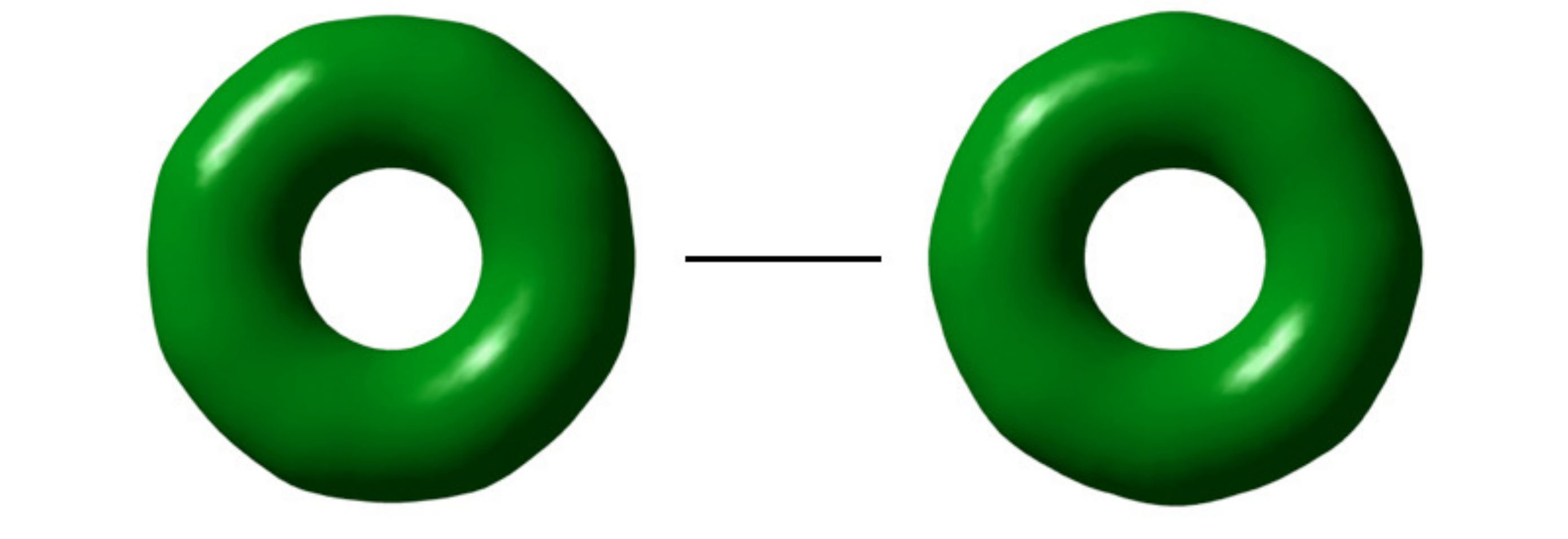}
		\label{fig:uhfcoreeq_a}
	}
	\subfloat[][core pair at 3 \AA]{
		\centering
		\includegraphics[width=2.9in]{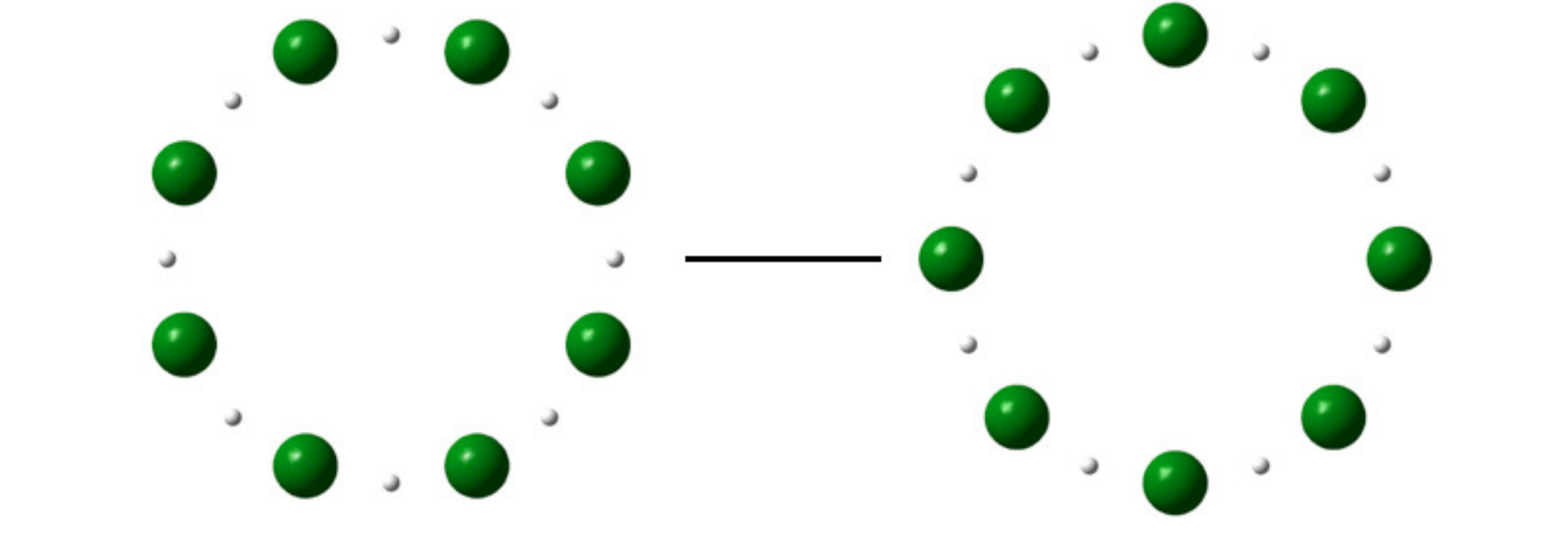}
		\label{fig:uhfcoreeq_b}
	}\\
	\subfloat[][HOMO pair at 1 \AA]{
		\centering
		\includegraphics[width=2.9in]{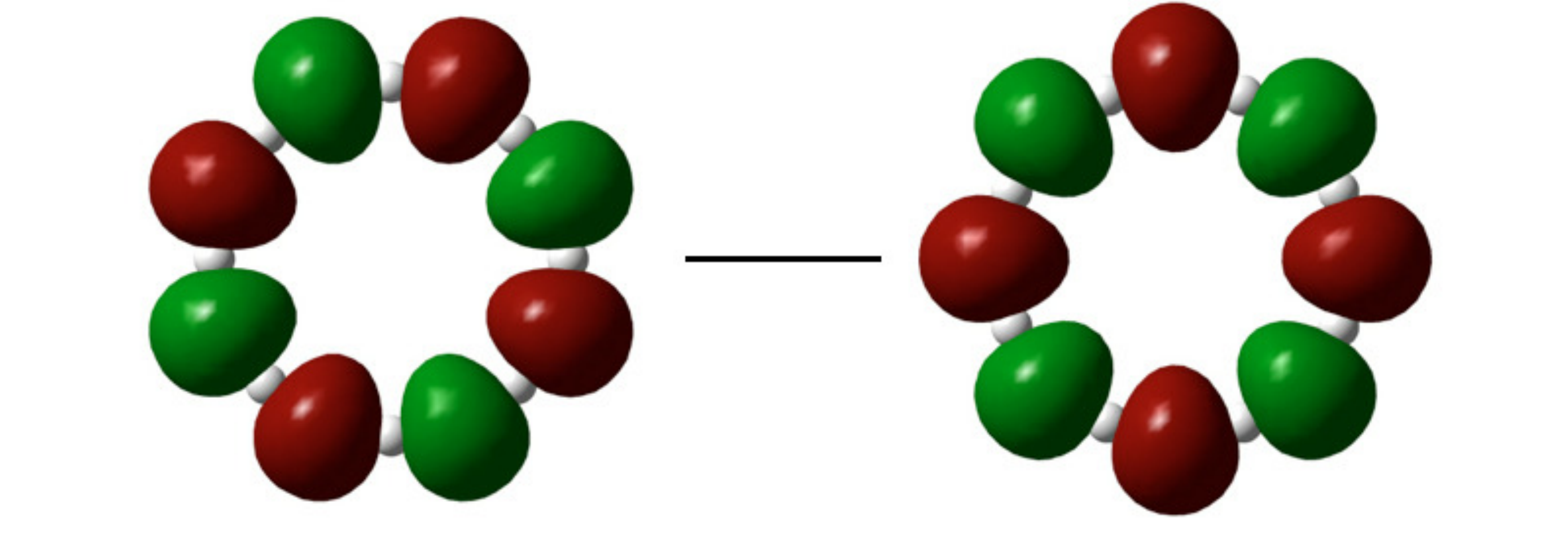}
		\label{fig:uhfhomoeq_a}
	}
	\subfloat[][HOMO pair at 3 \AA]{
		\centering
		\includegraphics[width=2.9in]{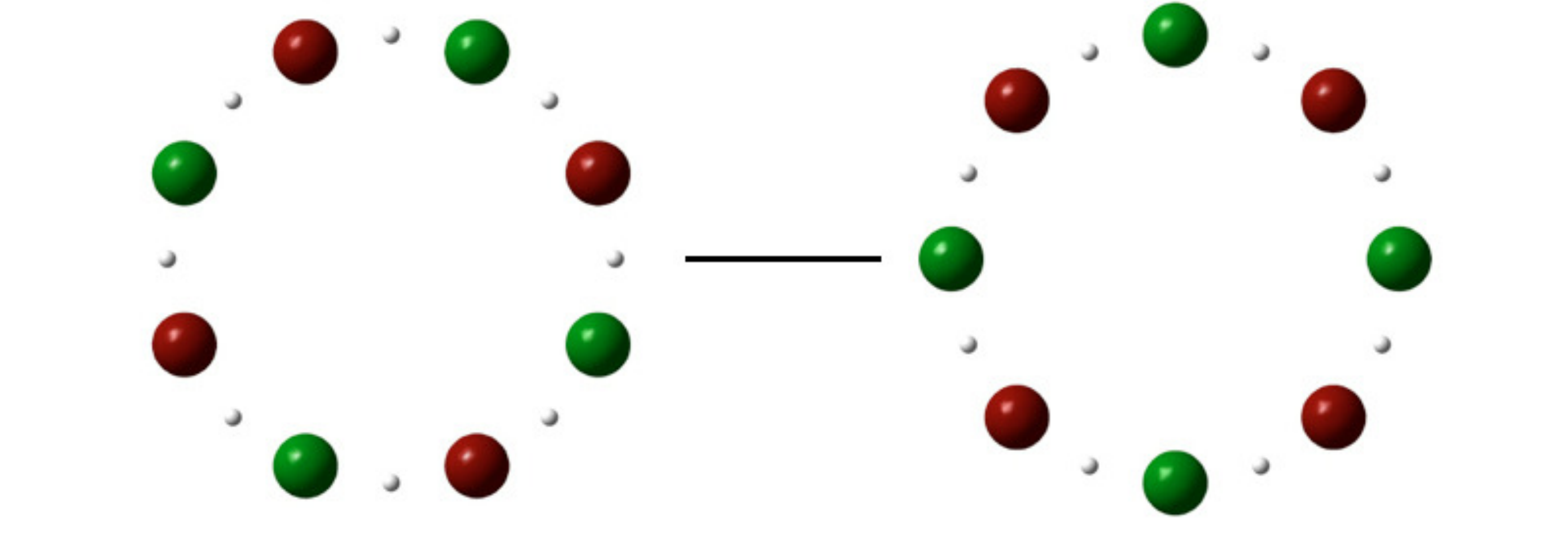}
		\label{fig:uhfhomoeq_b}
	}
	\caption[]{
	\label{fig:H16uhforbs}
	Core and HOMO UHF corresponding (pair) orbitals for the \ce{H16} ring. At large $R_{n-n}$ the broken symmetry wave function localizes the alpha and beta spins on different sublattices.
	}
\end{figure*}

Again, we pause to stress that the multi-reference SUHF wave function can be characterized by a single, deformed determinant, distinct from the self-consistent UHF wave function. To illustrate this point, we show the value of the overlap integral ($\sqrt{\xi}$) for core and degenerate HOMO orbitals for both the UHF and broken-symmetry determinant characterizing the SUHF wave function in the inset of Figure \ref{fig:h18-uhf-over}. By \textit{deliberately} breaking the symmetry of the underlying determinant in the SUHF approach, artificially localizing the electrons at small distances, and projecting the deformed determinant, the SUHF approach captures a significant amount of correlation energy near equilibrium.

Finally returning to the anti-resonant \ce{H16} ring, we plot the core and HOMO pair orbitals in Figure \ref{fig:H16uhforbs}. These are analogous to the case of H$_{4}$, and the HOMO orbital localizes one of each spin on separate sublattices at all distances. This creates an open-shell singlet, while the rest of the orbitals retain the freedom to localize or not as correlation effects demand. Near the CF point the core orbital remains delocalized, and by 3 \AA \ it localizes each spin onto different sublattices. The remaining orbitals behave in a similar manner.

\begin{figure*}[ht]
	\subfloat[][]{
		\centering
		\includegraphics[width=2.9in] {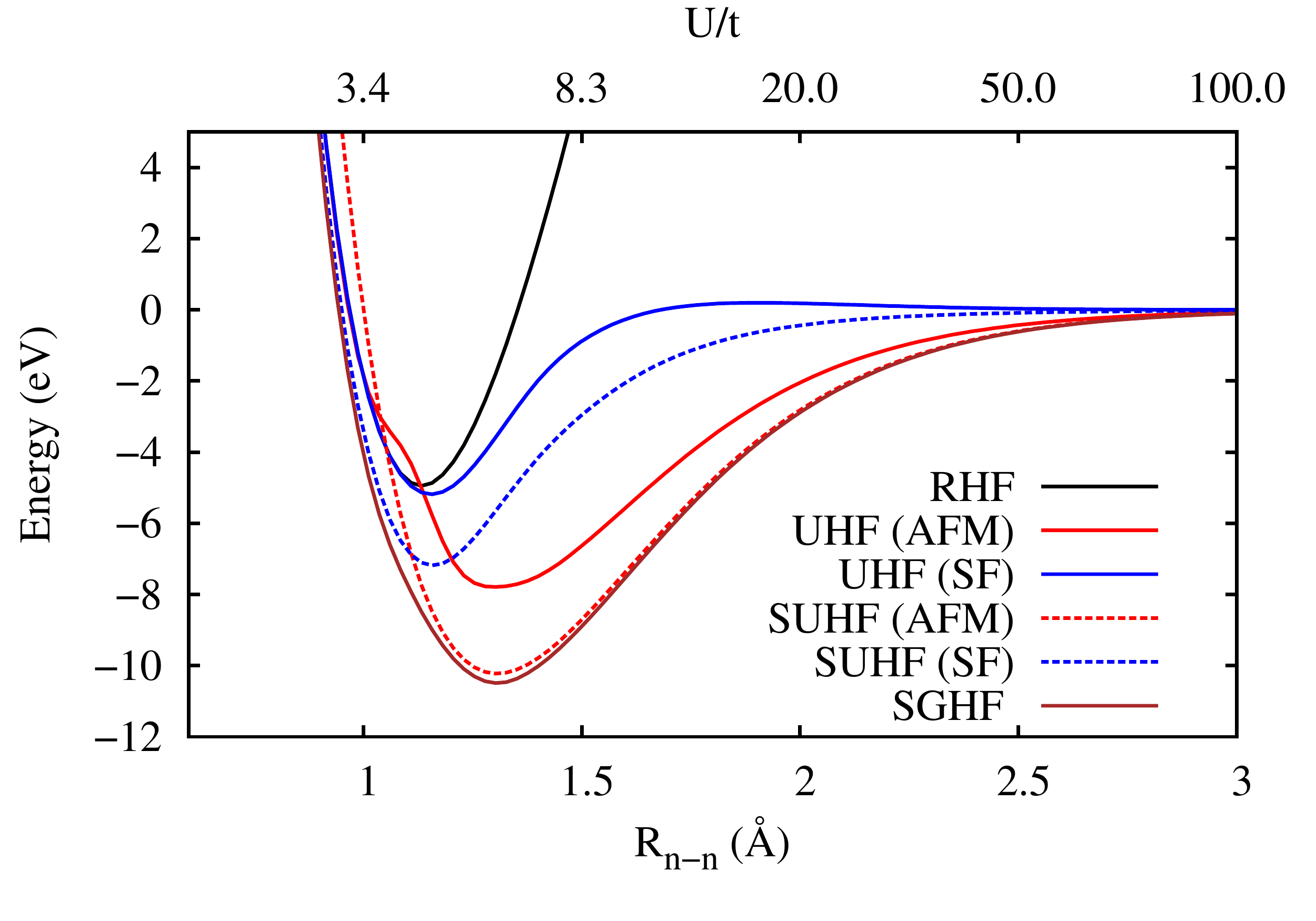}
		\label{fig:h16-square-dis}
	}
	\subfloat[][AFM Spin Density]{
		\centering
		\includegraphics[width=1.5in] {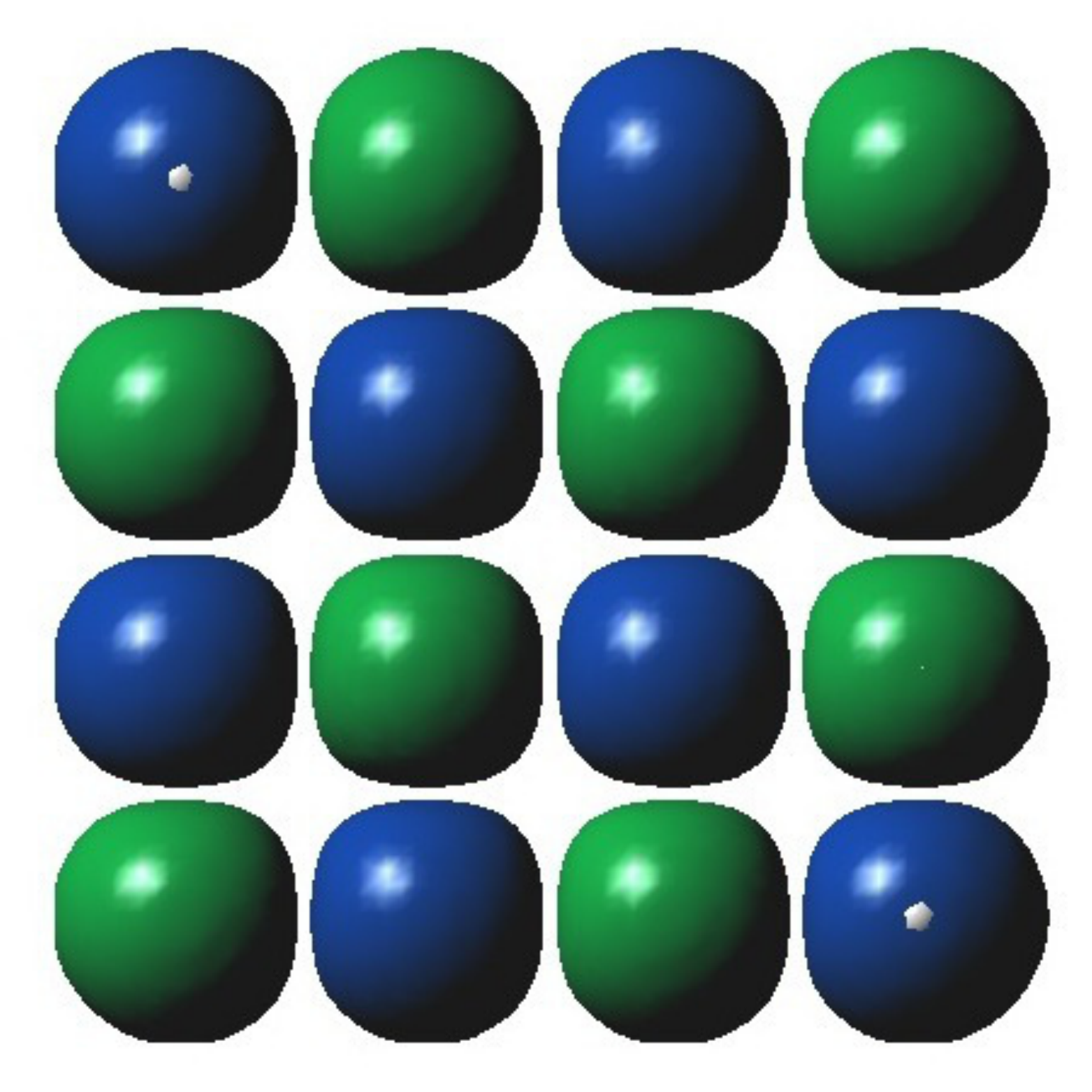}
		\label{fig:h16-s1}
	}
	\subfloat[][SF Spin Density]{
		\centering
		\includegraphics[width=1.5in] {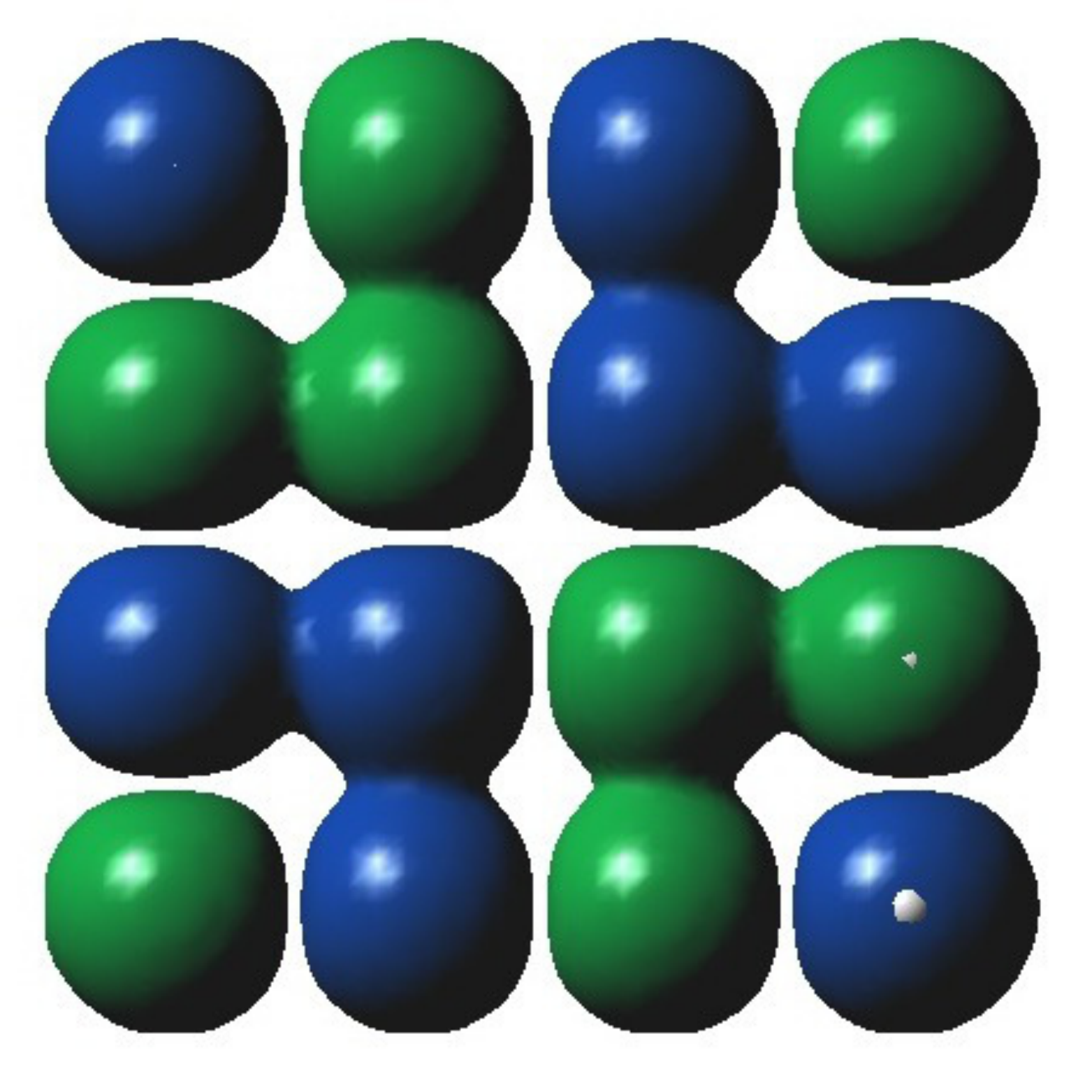}
		\label{fig:h16-s2}
	}\\
	\subfloat[][Pair 1  $\sqrt{\zeta}$=0.121]{
		\centering
		\includegraphics[width=1.5in] {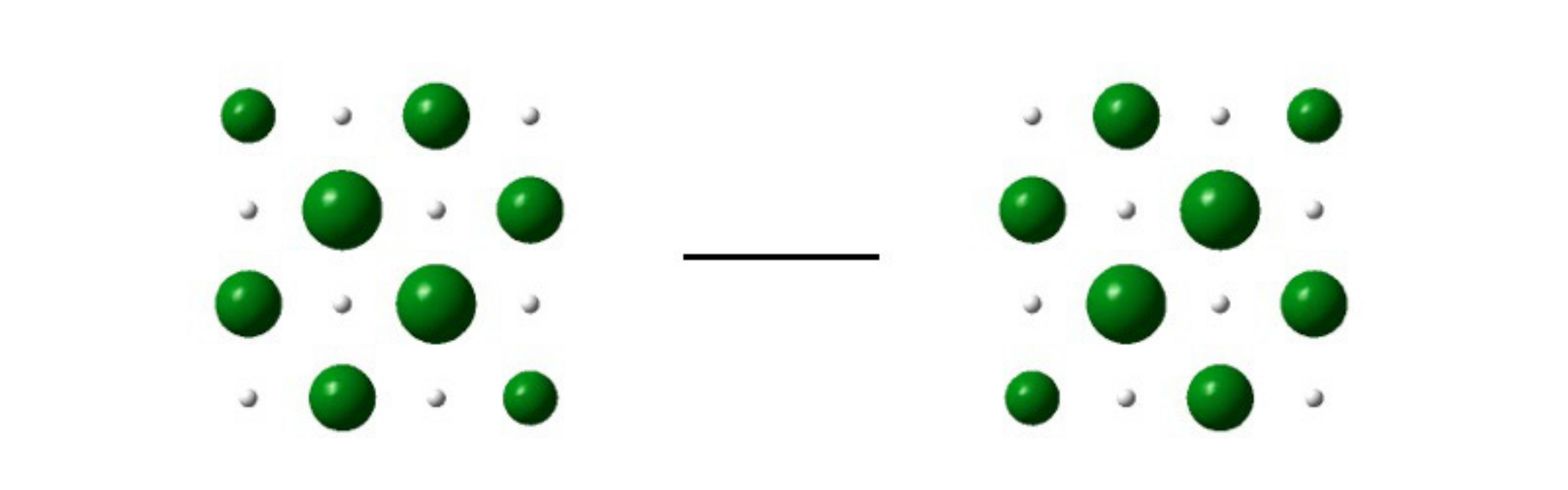}
		\label{fig:h16-p1}
	}
	\subfloat[][Pair 2  $\sqrt{\zeta}$=0.084]{
		\centering
		\includegraphics[width=1.5in] {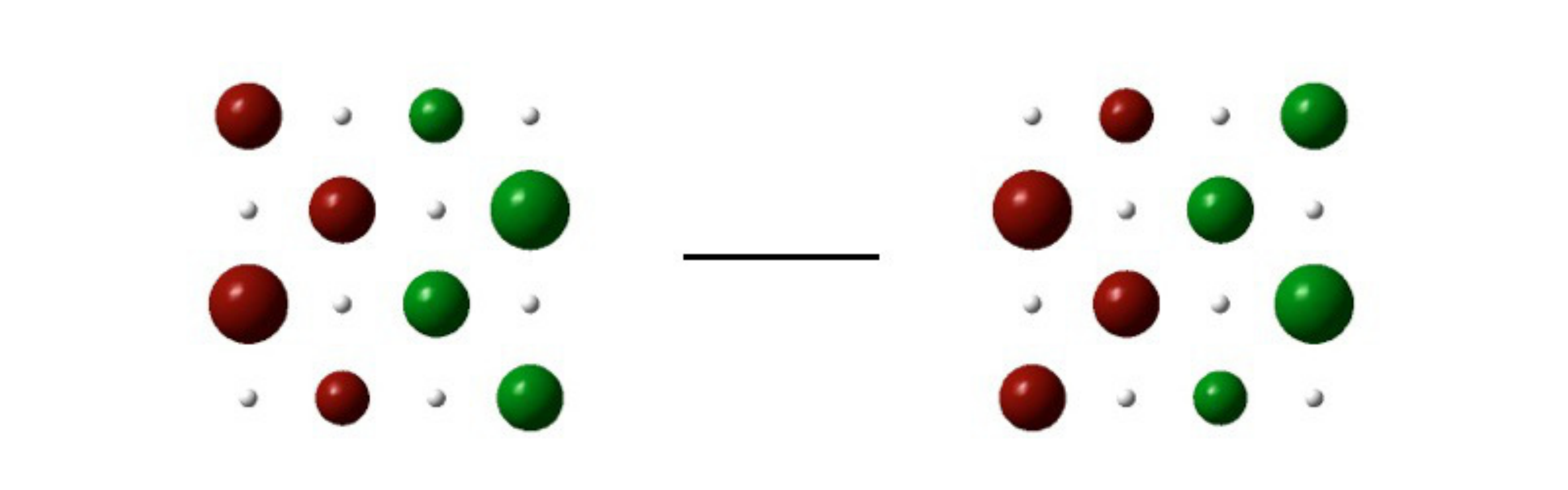}
	}
	\subfloat[][Pair 3  $\sqrt{\zeta}$=0.084]{
		\centering
		\includegraphics[width=1.5in] {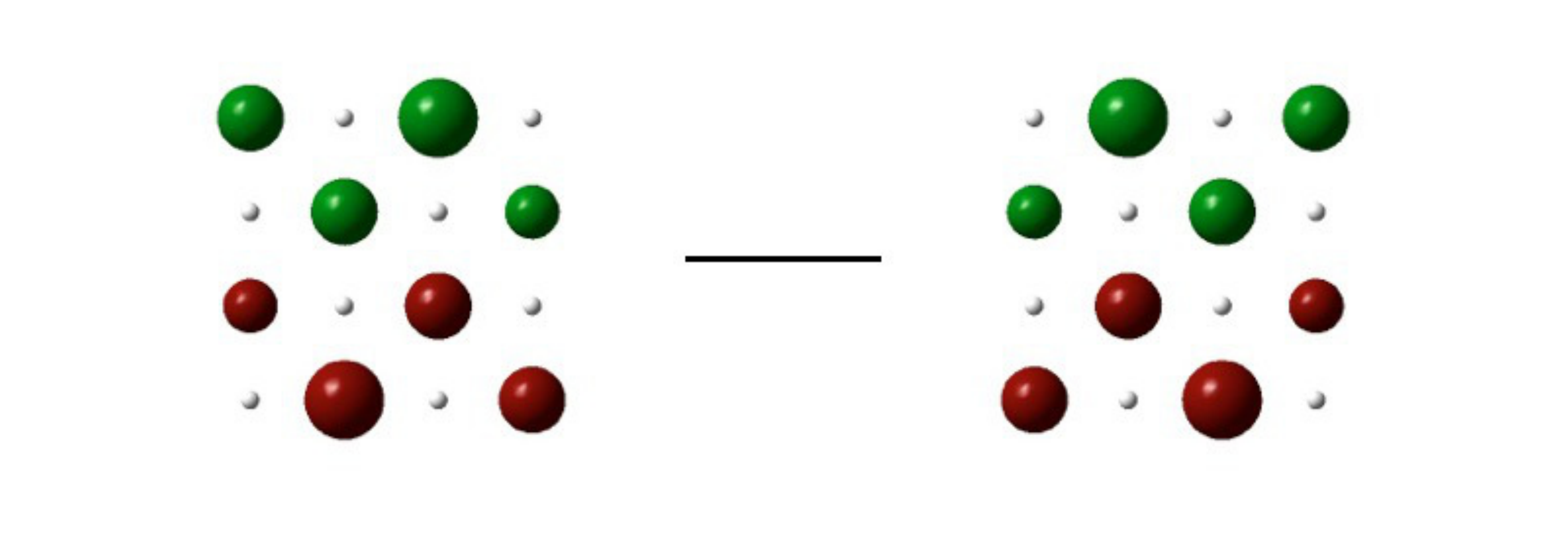}
	}
	\subfloat[][Pair 4  $\sqrt{\zeta}$=0.046]{
		\centering
		\includegraphics[width=1.5in] {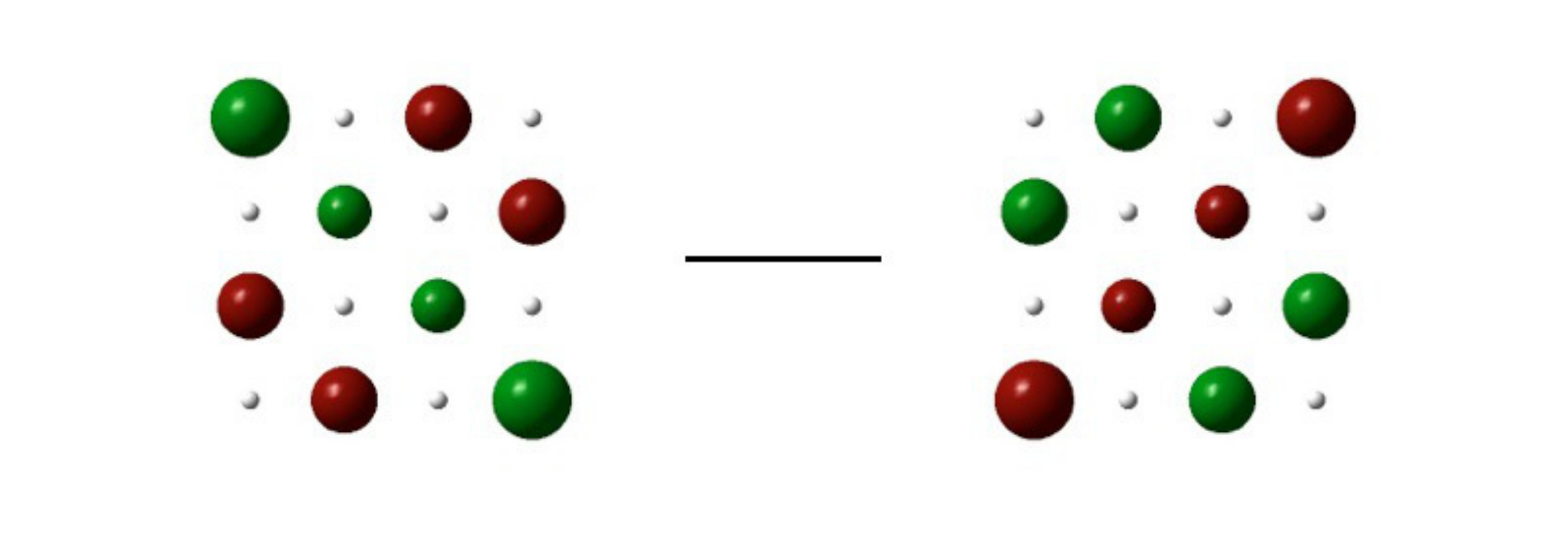}
	}\\
	\subfloat[][Pair 5  $\sqrt{\zeta}$=0.038]{
		\centering
		\includegraphics[width=1.5in] {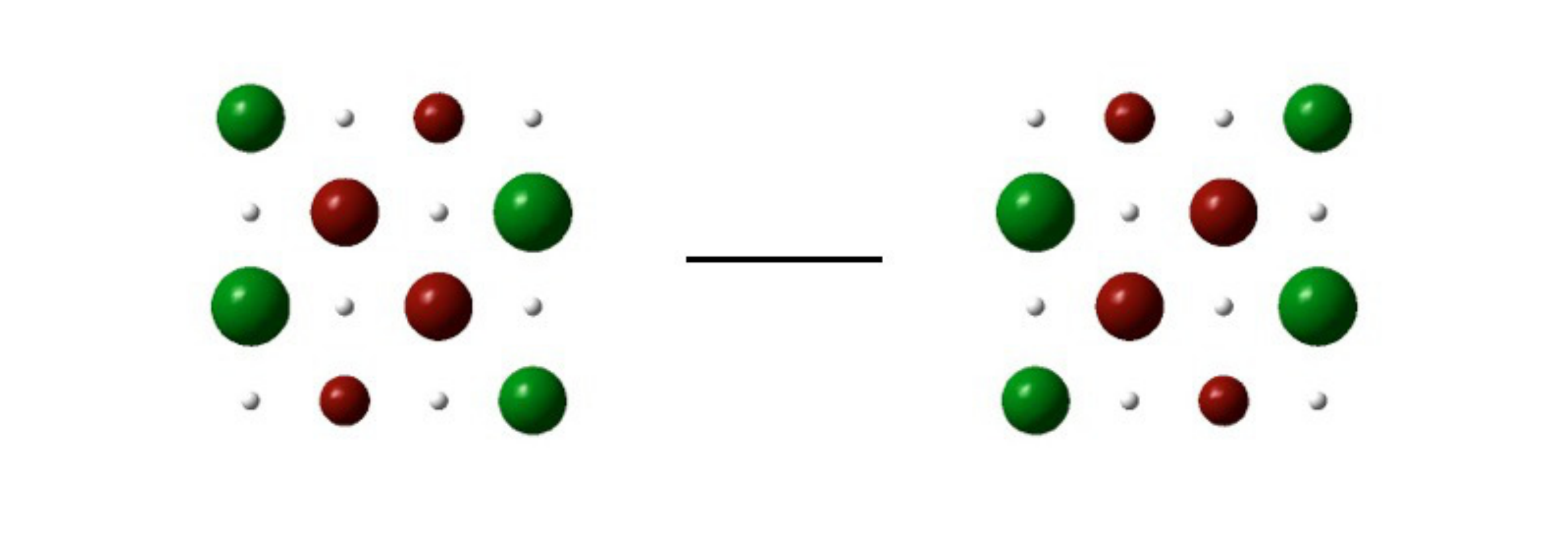}
	}
	\subfloat[][Pair 6  $\sqrt{\zeta}$=0.038]{
		\centering
		\includegraphics[width=1.5in] {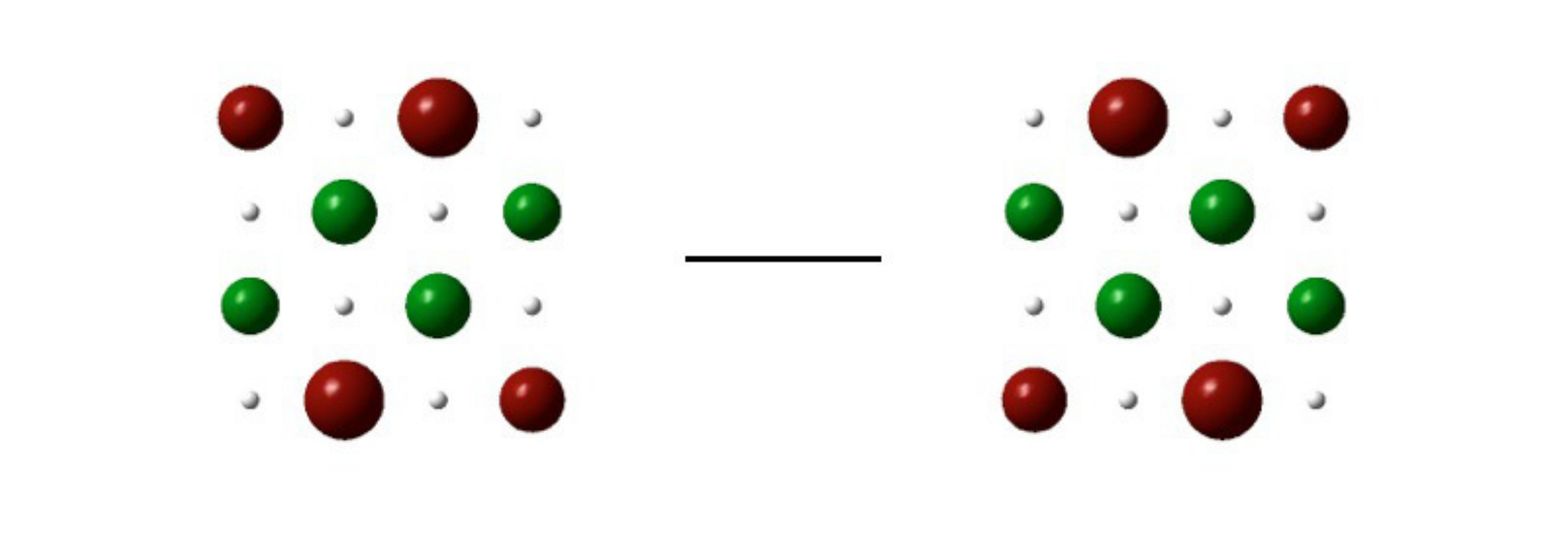}
	}
	\subfloat[][Pair 7  $\sqrt{\zeta}$=0.000]{
		\centering
		\includegraphics[width=1.5in] {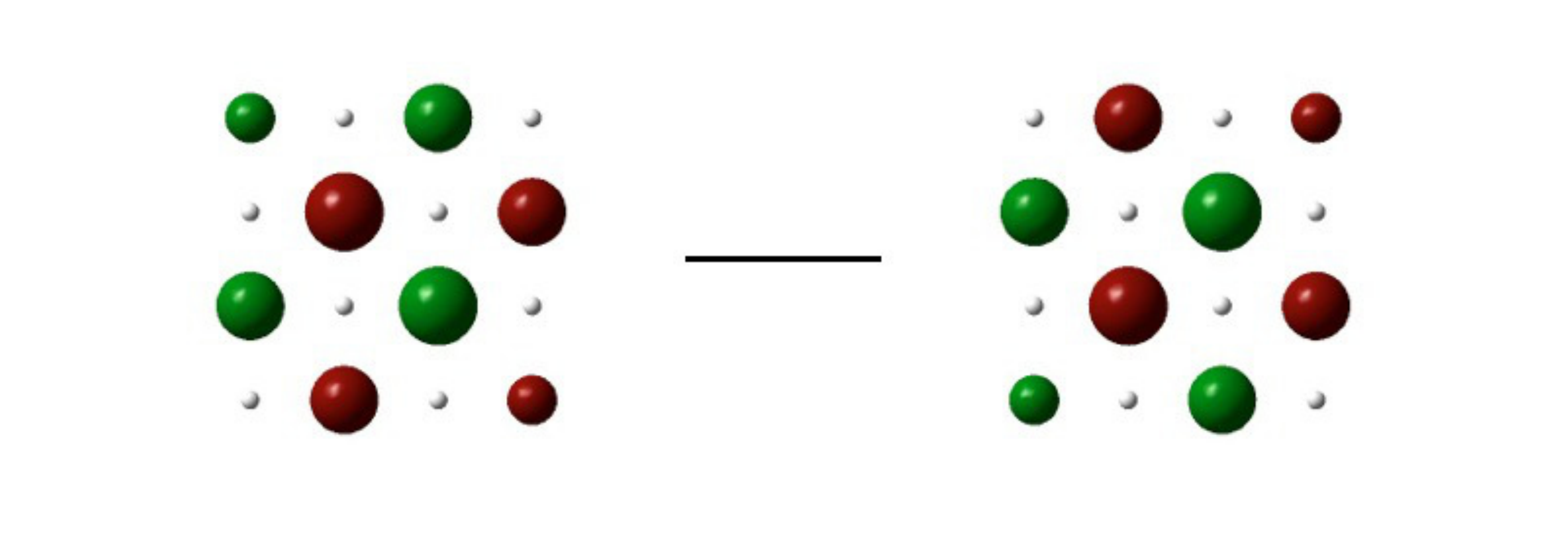}
	}
	\subfloat[][Pair 8  $\sqrt{\zeta}$=0.000]{
		\centering
		\includegraphics[width=1.5in] {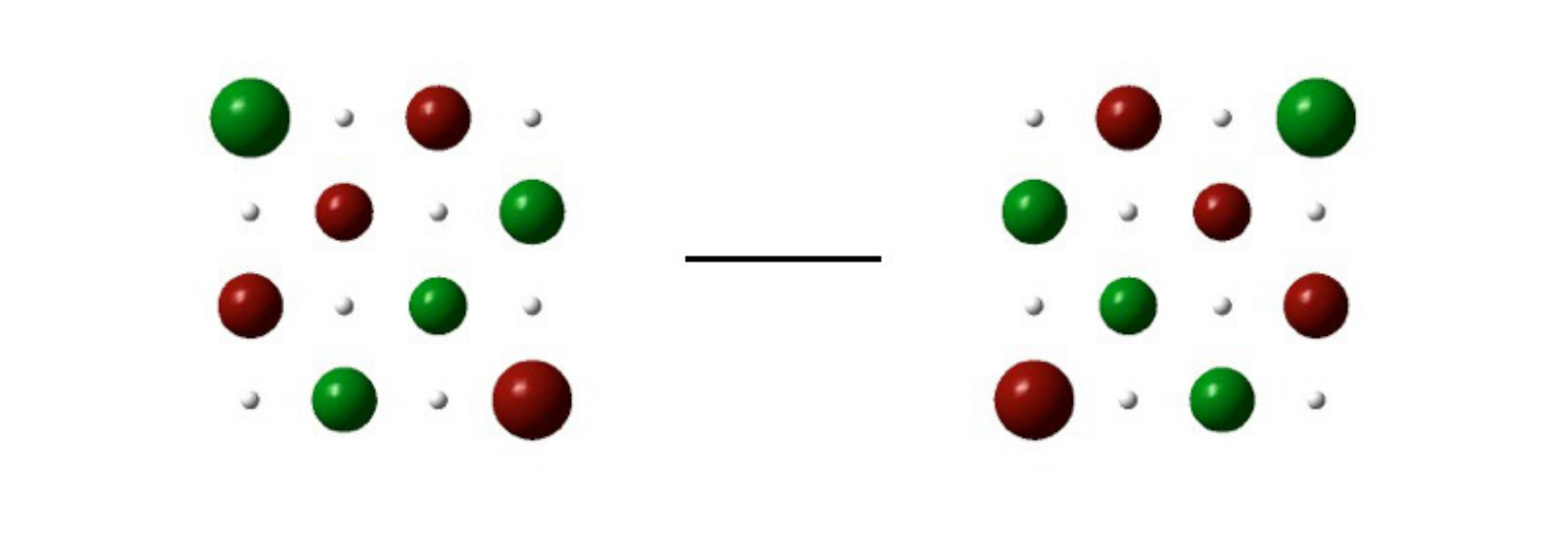}
		\label{fig:h16-p8}
	}
	\caption[]{
	\label{fig:H16square} The \ce{H16} plaquette.
	\subref{fig:h16-square-dis} Dissociation curves for singlet wave functions  \subref{fig:h16-s1} AFM state UHF spin density, and \subref{fig:h16-s2} SF state UHF spin density at $R_{n-n}$ = 3.0 \AA. 
\subref{fig:h16-p1}-\subref{fig:h16-p8} UHF pair orbitals and overlaps for the singlet AFM state at $R_{n-n}$ = 3.0 \AA.\bibnotemark[sym] At this distance, each pair localizes one alpha and one beta spin on different sublattices.
	}
\end{figure*}

\subsection{2D Hydrogen Networks}

We conclude with a brief discussion of a 2D system: symmetric dissociation of \ce{H16} in a square geometry. The square lattice has been previously studied at half-filling by Larson and Thorson, and Calais and coworkers.  \cite{1966JChPh..45.1539L,Calais:1965wl,Calais:1965tx,Calais:1965vi,Calais:1965ua} Though the  system studied here doesn't have periodic boundary conditions, it still captures the essential Hubbard physics. We point out that the UHF single particle energy spectrum differs from that of the \ce{H16} ring in that the degeneracy of the HOMO has lifted and the Fermi energy now lies within the gap. This results in a close-shell singlet ground state, which becomes the focus of our discussion.

Interestingly, we find two distinct stable UHF states, shown in Figure \ref{fig:h16-square-dis}. The difference between these two states is subtle, amounting to a ninety degree rotation of the localized spins in the center of the plaquette (see Figures  \ref{fig:h16-s1} and \ref{fig:h16-s2}). We also plot the pair orbitals for the AFM state near dissociation,\bibnotemark[sym] showing the segregation of respective spins onto separate sublattices. The UHF global minimum energy occurs for the AFM state, however, the orbital Hessian indicates that the SF state exists as a true local minimum, and not a transition state. One might naively hope that projecting these two different UHF singlet states would yield the same multi-reference wave function for the singlet state. Unfortunately, this intuition fails for the SUHF approach. 

As mentioned previously, the SCF procedure preserves the symmetries of the initial guess for the broken symmetry determinant; consequently, we find two distinct SUHF singlets. Using the SGHF approach overcomes this problem: breaking $\hat{S}_{z}$ allows the dissociation curve to smoothly interpolate between these two different spin densities. The SGHF wave function captures the most correlation in the region where these two states become energetically degenerate, as one would expect. Note that we have not performed a projection to restore spatial symmetry; the SGHF wave function only restores $\hat{S}^{2}$ and $\hat{S}_{z}$. 

We also find multiple UHF and SUHF solutions arising from broken spatial symmetries in larger models such as 8x8 plaquettes. One can in some sense think of these different states as incorporating different types of spin fluctuations. By interpolating between these spin states, the SGHF approach captures additional correlation in the region where states of different spatial symmetry become energetically degenerate. Ongoing work indicates that breaking and restoring point group symmetry often becomes very important in cases such as these where the SGHF wave function does not have enough flexibility to describe all of the static correlation in the system.

We conclude this section by pointing out that we may apply the pairing ansatz which segregates spins onto separate sublattices to any alternant lattice: the one-dimensional, square, simple cubic and bcc lattices. For example: in terms of the Bloch functions of the original 2D lattice, the pairs mix the configurations $\psi_{kx,ky}^2$ with their anti-bonding partner $\psi_{kx+\pi,ky+\pi}^2$. The wave function may be written as a spin-contaminated APSG if the corresponding pairs at each k-value come from the corresponding orbitals of the spin-polarized solution. The energy of this wave function is identical with that of the N\'{e}el state. In fact, the energies of all the pure spin states are degenerate, and the N\'{e}el state a linear combination of them.

\section{Conclusion}
\label{sec:Conclusion}

We have emphasized that the UHF wave function can be written as a spin-contaminated \textit{pair} wave function of the APSG form. The overlap of the alpha and beta corresponding orbitals of the UHF solution can be taken as a proxy for the strength of the correlation captured by breaking symmetry.  As a function of distance, or the ratio U/t, the UHF corresponding orbitals pair in a manner allowing a smooth evolution from a system with doubly occupied orbitals into one in which the $\alpha$ and $\beta$ electrons segregate onto distinct sublattices. In this way the UHF wave function evolves from a regime in which the material behaves as a metal with delocalized spins, to one in which all the spins are localized antiferromagnetically.  

Projecting the spin-contaminated UHF pair wave function recovers additional correlation energy in finite systems. The resulting multi-reference wave function, characterized by a single, deformed determinant, describes a pure spin state. The SUHF wave function (projecting $\hat{S}^{2}$ by assuring rotational invariance in spin-space) captures additional intra-pair correlation beyond UHF. The SGHF wave function (projecting $\hat{S}^{2}$ and $\hat{S}_{z}$) captures additional inter-pair correlation. By \textit{deliberately} breaking and then restoring these symmetries, we describe strong correlation even at geometries where the UHF solution reduces to RHF or the GHF solution reduces to UHF. 

From simple calculations of one and two-dimensional finite lattices of hydrogen atoms, we have shown that the broken symmetry determinant characterizing our SUHF solutions mirror the pairing schemes utilized extensively some years ago in AMO theory to describe correlation in alternant pi-bonded networks. This early work did not address the questions of what happens as you dope the system away from half-filling, or in what manner the pairs respond. The pairs described here accounting for magnetic correlations are strongly orthogonal to one another, whereas the pairs of BCS theory overlap; nevertheless, the proximity of superconducting behavior to antiferromagnetic phases suggest this may be an interesting avenue to explore. 

\acknowledgement
Our work at Los Alamos National Laboratory was supported by the Department of Energy, Office of Basic Energy Sciences, Heavy Element Chemistry program  and the LDRD program at LANL. 
The work at Rice University was supported by DOE, Office of Basic Energy Sciences, Heavy Element Chemistry program under Grant DE-FG02-04ER15523. The Los Alamos National Laboratory is operated by Los Alamos National Security, LLC, for the National Nuclear Security Administration of the U.S. Department of Energy under Contract DE-AC5206NA25396.

%



\begin{mcitethebibliography}{49}
\bibitem[sym()]{sym}
The perfect geometric symmetry of the ring was broken slightly by adding a
  small ($0.01$\ \r A) displacement to each succeeding atom along the vertical
  z-axis. This imposed helical distortion allows us to uniquely define each of
  the molecular orbitals of the system, as it lifts the perfect degeneracy in
  the $e$ symmetries, and removes the freedom of the diagonalization routine to
  rotate them in an arbritrary manner.\relax
\mciteBstWouldAddEndPunctfalse
\mciteSetBstMidEndSepPunct{\mcitedefaultmidpunct}
{}{\mcitedefaultseppunct}\relax
\EndOfBibitem
\end{mcitethebibliography}

\providecommand*\mcitethebibliography{\thebibliography}
\csname @ifundefined\endcsname{endmcitethebibliography}
  {\let\endmcitethebibliography\endthebibliography}{}

\end{document}